\documentclass[aip,jap,preprint]{revtex4-1}
\usepackage{mathtools}
\usepackage{graphicx}
\usepackage{color}
\usepackage{bm}
\usepackage{dcolumn}
\usepackage{comment}
\usepackage{subcaption}
\usepackage{epstopdf}
\usepackage[normalem]{ulem}
\usepackage{amsmath}
\usepackage[font=sf]{caption}
\usepackage{relsize}
\usepackage[strings]{underscore}
\usepackage{float}
\usepackage{textcomp}
\usepackage{array}
\usepackage{titlesec}
\usepackage[font=sf]{caption}
\usepackage{multirow}

\newcommand{\dOUT}{\bgroup\markoverwith{\textcolor{blue}{\rule[.4ex]{1.pt}{1.5pt}}}\ULon}   
\usepackage{xcolor}

\AtBeginDocument{}
\begin{document}
\title{Ensemble Monte Carlo for III-V and Si n-channel FinFETs considering non-equilibrium degenerate statistics and quantum-confined scattering}
\author{Dax M. Crum}
\email{dcrum@utexas.edu}
\affiliation{Microelectronics Research Center, The University of Texas at Austin \\10100 Burnet Road, Austin, Texas 78758, USA}
\author{Amithraj Valsaraj}
\affiliation{Microelectronics Research Center, The University of Texas at Austin \\10100 Burnet Road, Austin, Texas 78758, USA}
\author{John K. David}
\affiliation{Microelectronics Research Center, The University of Texas at Austin \\10100 Burnet Road, Austin, Texas 78758, USA}
\affiliation{Currently with Intel Corporation \\2501 Northwest 229th Ave., Hillsboro, Oregon 97124, USA}
\author{Leonard F. Register}
\affiliation{Microelectronics Research Center, The University of Texas at Austin \\10100 Burnet Road, Austin, Texas 78758, USA}
\author{Sanjay K. Banerjee}
\affiliation{Microelectronics Research Center, The University of Texas at Austin \\10100 Burnet Road, Austin, Texas 78758, USA}
\date{\today}
%
%
\begin{abstract}
Particle-based ensemble semi-classical Monte Carlo (MC) methods employ quantum corrections (QCs) to address quantum confinement and degenerate carrier populations to model tomorrow's ultra-scaled MOSFETs.  Here we present the most complete treatment of quantum confinement and carrier degeneracy effects in a three-dimensional (3D) MC device simulator to date, and illustrate their significance through simulation of n-channel Si and III-V FinFETs. Original contributions include our treatment of far-from-equilibrium degenerate statistics and QC-based modeling of surface-roughness scattering, as well as considering quantum-confined phonon and impurity scattering in 3D. Typical MC simulations approximate degenerate carrier populations as Fermi distributions to model the Pauli-blocking (PB) of scattering to occupied final states. To allow for increasingly far-from-equilibrium non-Fermi carrier distributions in ultra-scaled and III-V devices, we instead generate the final-state occupation probabilities used for PB by sampling the local carrier populations as a function of energy and energy valley. This process is aided by the use of fractional carriers or sub-carriers, which minimizes classical carrier-carrier scattering. Quantum confinement effects are addressed through quantum-correction potentials (QCPs) generated from coupled Schr{\"o}dinger-Poisson solvers, as commonly done. However, we use our valley- and orientation-dependent QCPs not just to redistribute carriers in real space, or even among energy valleys, but also to calculate confinement-dependent phonon, impurity, and surface-roughness scattering rates. FinFET simulations are used to illustrate the contributions of each of these QCs. Collectively, these quantum effects can substantially reduce and even eliminate otherwise expected benefits of considered In$_{\text{0.53}}$Ga$_{\text{0.47}}$As FinFETs over otherwise identical Si FinFETs, despite higher thermal velocities in In$_{\text{0.53}}$Ga$_{\text{0.47}}$As.
\end{abstract}
\maketitle
\section{Introduction}
Multi-gate metal-oxide-semiconductor-field-effect-transistors (MOSFETs) have supplanted planar MOSFETs as the clear device choice for future integrated circuit technology. The three-dimensional (3D) fin-shaped MOSFET, or FinFET,~\cite{hisamoto} is electrostatically superior~\cite{sze,rado} to its planar relatives and already directing current technologies and future complimentary MOS (CMOS) scaling.~\cite{ramey,ang} In addition, high mobility III-Vs are being considered as channel replacements for Si.~\cite{dewey,alamo} In$_{\text{0.53}}$Ga$_{\text{0.47}}$As, which is lattice-matched to fabrication-friendly InP,~\cite{mukherjee} is being considered to provide a drive-current boost via light-effective mass carriers with associated large thermal injection velocities.~\cite{lee,rado2} 

Modeling such devices presents challenges for predictive device simulators, which are needed to optimize the large design space and estimate future scaling benefits. Alternate channel materials and associated transport physics require a microscopic description of their behavior. Quasi-ballistic transport cannot be completely described by continuum diffusive models in these devices,~\cite{seoane, choi, L3} yet scattering remains crucial, even as channel lengths are scaled well below 22 nm.\cite{L1,L2,sviz} Certainly fully coherent methods, such as non-equilibrium Green function (NEGF) techniques,\cite{datta} have demonstrated their value to studying such systems~\cite{martinez,amoroso,kmliu,chen} and represent the reference standard in the ballistic limit. However, upon the inclusion of scattering in realistic device geometries, pure quantum methods can become computationally impractical for many applications. Non-randomizing polar optical phonon scattering, which dominates $\Gamma$-valley transport in III-V channels as considered here, still has not been achieved in 3D NEGF simulations. Therefore it still remains important to extend the validity of semi-classical methods via so-called quantum corrections (QCs) to model these nanoscale devices while maintaining reasonable computational efficiency.

For these reasons, particle-based ensemble semi-classical Monte Carlo (MC) remains a benchmark in semiconductor device research.  It allows modeling of various distinct scattering mechanisms (including non-randomizing processes) and consideration of complex device geometries. MC is known to predictively model diffusive through ballistic transport including non-local field effects such as velocity overshoot.  MC, however, suffers from its own drawbacks. Traditional particle-based MC is rigorous only in large systems where the carrier distributions are well-approximated by the bulk energy dispersion relations and scattering rates. Cutting-edge electron devices, however, often go well beyond these limits. In today's maximally-doped source and drain (S/D) transistor reservoirs and above threshold in the channel, degenerate carrier populations must be considered, along with associated Pauli-blocking (PB) of scattering.  FinFET fin widths of a few nm (already 8~nm in 22~nm node devices~\cite{ramey}) modify not only the carrier distributions in real-space but also the band structure and even scattering rates. Each of these quantum effects is exacerbated in materials with very light effective masses $m^*$ (e.g., $m^* = 0.044~m_{\mathrm{e}}$ for $\Gamma$-valley electrons in In$_{\text{0.53}}$Ga$_{\text{0.47}}$As), now being considered for MOSFETs. The continued use of particle-based MC under these conditions requires modification to the semi-classical methodology.  

In this work, we present an ensemble 3D semi-classical MC simulator for n-channel devices whose treatment of electron degeneracy and quantum confinement institutes new approaches for particle-based simulations. We focus on our original contributions to the state-of-the-art, including our treatment of (i) far-from-equilibrium degenerate statistics, (ii) QC-based modeling of surface-roughness scattering, and (iii) extending our group's previously introduced treatment of quantum-confined phonon and impurity scattering to 3D. In doing so, we expand upon our techniques, verify our methodologies, and refine and extend results introduced in a short preliminary study.~\cite{crum}

After a brief description of the underlying purely semi-classical simulator, we detail our treatment of far-from-equilibrium degenerate carrier statistics. To consider the Pauli exclusion principle in MC simulation, scattering processes for electrons are either accepted or rejected according to the probability that the final scattering state is already occupied. The distributions of final states are typically approximated as being Fermi distributions, even if hot, dictated by the average local electron density and energy.~\cite{laux2,pantoja,mateos,islam,kalnaFD,islam2,david} However, this approximation cannot be justified under strong non-equilibrium conditions approaching the ballistic limit of performance. In this work, we avoid \textit{a priori} assumptions about the shape of the electronic distribution functions. Instead, we sample the electron populations locally in energy, energy valley, and propagation direction to generate the occupation numbers for the PB of scattering to states which are already occupied. Such approaches have been executed in $\mathbf{k}$-space for bulk calculations,~\cite{jacoboni, jacoboni2} but now we extend this method to include real-space variations in the distribution function for device simulation.  This process is aided by the use of fractional carriers or sub-carriers, which not only improves statistics but, as the principle motivation, minimizes classical carrier-carrier scattering otherwise introduced via the time-dependent solution of Poisson's equation, which is incompatible with degenerate statistics.     

Next, QCs for various quantum-confinement effects are provided through multiple uses of valley-, space-, orientation-, and time-dependent quantum-correction potentials (QCPs). Here, we calculate the set of QCPs based on the solutions of effective mass Schr{\"o}dinger's equations defined in each channel slice normal to the transport direction~\cite{winstead,katha,fan,reg1,shi,shi2,*[{}] [{ Ph.D. dissertation, The University of Texas at Austin.}] david2,david3,lindberg,nagy} on a valley-by-valley basis~\cite{reg1,shi,shi2,david2,david3} considering 2D confinement,~\cite{david2,david3,lindberg,nagy} a first-principles strategy requiring no adjustable parameters.  However, it is our uses of the QCPs, not their method of calculation, which is the focus here. Indeed, it may be possible to extend such uses of QCPs in these ways, however calculated, to still more computationally efficient drift diffusion and hydrodynamic simulations.  The QCPs redistribute the MC electrons in real-space (e.g., away from interfaces) to reflect the quantum-mechanical spatial density. In addition, our QCPs naturally alter energy separations between energy valley minima, leading to degeneracy-splitting and redistribution of charge among energy valleys through scattering. Further, we use the QCPs to adjust 3D phonon and ionized-impurity scattering rates self-consistently on-the-fly, an extension of our previous 2D strategy.~\cite{reg1,shi,shi2} Lastly, for the first time in MC simulation, we model surface-roughness (SR) scattering rates as a function of our QCPs. This SR method is quite general and allows for arbitrary potential-well shapes and confining geometries, moving beyond typical triangular-well assumptions for SR rate calculations.

Our QCs capture the main qualitative effects of quantum confinement and electron degeneracy within MC simulation. Using In$_{\text{0.53}}$Ga$_{\text{0.47}}$As and Si n-channel FinFETs as examples, we illustrate the importance of each quantum effect by analyzing simulation results with and without QCs. While both III-V and Si devices suffer these quantum effects, the scale is decidedly greater for III-V devices. In In$_{\text{0.53}}$Ga$_{\text{0.47}}$As $\Gamma$-valleys, the light effective masses, low densities of states (DOS), and encountered highly-degenerate carrier populations lead to undesirable low quantum capacitances and high SR scattering rates. Confinement-reduced intervalley energy separations lead to sizable transfer of $\Gamma$-electrons to peripheral L- and X-states, reducing channel injection velocities, although also beneficially increasing the quantum capacitance. A measure of the significance of the here-modeled quantum effects is that the simulated ultra-scaled III-V devices exhibit worse ON-state transconductance than otherwise identical Si devices.

In Section II, we introduce our device simulator before discussing the details of our QCs in Sections III and IV. In Section V, we present a detailed comparison of devices with different levels of quantum-corrected modeling. Finally, we summarize our study in Section VI.
\section{3D FinFET device and uncorrected MC simulator}
The device structure and the baseline purely semi-classical MC simulator used in this work are intended as vehicles for illustrating the QCs that are the focus of this study. Indeed, both the structure and baseline MC simulator are somewhat idealized for this purpose.  
\subsection{Device structure}
\begin{figure}[t]
	\includegraphics[width=0.95\columnwidth]{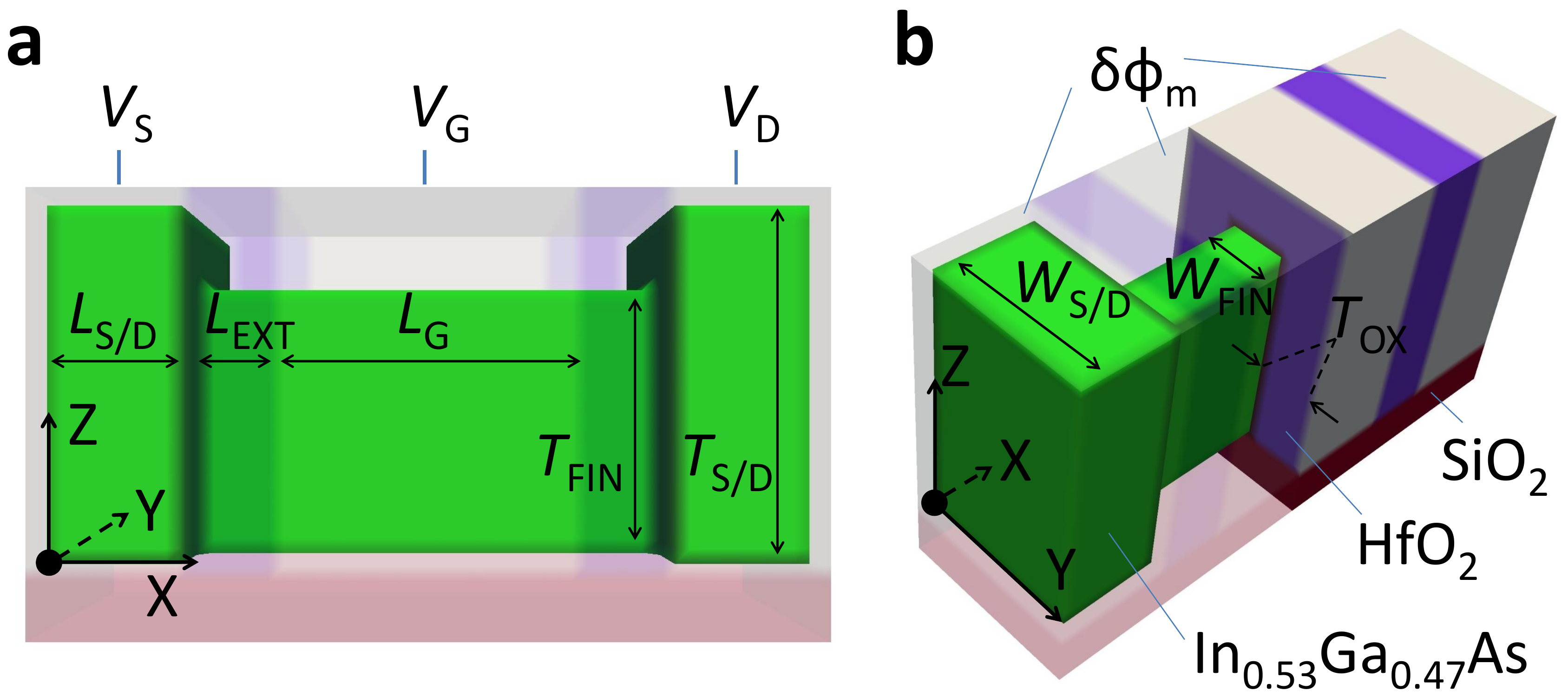}
	\caption{\label{fig:figFIN} \textbf{\textbar\; FinFET structure. (a)} Side view of the simulated device with relevant dimensions. Outer materials are shown transparent to visualize the semiconductor fin. \textbf{(b)} Edge view with cross-sectional clip to show different material regions. We vary the work functions in the $\delta\Phi_{\mathrm{m}}$ metal contacts to set the injection boundary conditions in the S/D such that there are flat-band conditions at the interface.  }
\end{figure}

The device used as a test bed in this work is shown in Fig.~\ref{fig:figFIN}.  It has a fin-shaped semiconducting channel, connecting two heavily-doped electron reservoirs. The n-type channel materials we study here are industry-standard Si and In$_{\text{0.53}}$Ga$_{\text{0.47}}$As, the latter being lattice-matched to fabrication-friendly InP~\cite{mukherjee} and a candidate for future CMOS. The S/D reservoirs are doped to $N_{\mathrm{D}} = 5\times10^{19}~\mathrm{cm}^{-3}$, a realistic activated dopant density which can be reached in III-V materials with current \textit{in situ} growth technology.~\cite{charache} Certainly Si devices are doped much more heavily than this in practice, as in our separate ongoing simulation study focused on device scaling and short-channel performance. Here, however, the focus is on our simulation methods, so we choose equal doping concentrations between the materials as a control. This allows a more fair comparison with regard to the essential transport physics.  We also consider a lower dopant density of $1\times10^{19}~\mathrm{cm}^{-3}$ in In$_{\text{0.53}}$Ga$_{\text{0.47}}$As. The correspondingly lower chemical potential in the S/D avoids contact injection directly into the peripheral valleys in In$_{\text{0.53}}$Ga$_{\text{0.47}}$As, isolating the role of intervalley scattering within the device simulation region.  We model all devices as having perfectly injecting and absorbing boundary conditions, an idealization for both systems but more so for In$_{\text{0.53}}$Ga$_{\text{0.47}}$As, who requires more careful materials processing than Si to develop ohmic contacts.~\cite{singi} Perfectly injecting and absorbing boundary conditions give a more fair comparison between the channel materials by decoupling the channel performance from current experimental and technological constraints regarding the metal contacts. The undoped fin-channel sits atop a 5~nm insulating layer of SiO$_2$ ($\varepsilon_{\mathrm{r}} = 3.9$) in a semiconductor-on-insulator (SOI) fully-depleted configuration. The gate oxide is a 5~nm insulating layer of HfO$_2$ ($\varepsilon_{\mathrm{r}} = 22.3$) and is wrapped around a 6~nm wide fin-channel. The channel is 20~nm long beneath the gate with 6~nm extensions. The channel sidewall orientation is (surface)/$\langle$channel$\rangle$ = (100)/$\langle$100$\rangle$, which is optimized for n-type transport in Si, although not for Si CMOS as a whole. This orientation is more interesting in terms of the quantum effects due to the nature of the valley degeneracy-breaking, as will be discussed later.
\subsection{Baseline purely semi-classical Monte Carlo simulator}
Our baseline MC simulator follows the basic methods described in Refs.~\onlinecite{jacoboni,jacoboni2}. In later sections and plots, this purely semi-classical model (CL) provides a reference. It contains no considerations for the Pauli exclusion principle or quantum-confinement effects. We summarize important details of the implementation here. Specifically, we generate a uniform 3D real-space cubic mesh of 1~nm resolution. Each MC loop has a 1.2~fs time step such that, e.g., a very fast carrier moving $8\times10^{7}$~cm/sec can almost cross one grid site per time step. This time length is chosen large enough to minimize the computational burden while being small enough to converge the numerical data.  Poisson's equation is solved every time step consistent with the updated electron density. The electrostatics are modeled for each material based on their static dielectric permittivity.  Simulated electrons couple to the mesh via a nearest grid-point assignment of charge.  This approach is simple and eliminates self-forces while any electron remains in the cubic~nm cell centered about its grid site. However, when carriers do cross a cell boundary between grid points during a time step, the forces and scattering rates are adjusted instantaneously.  

At this point a significant self-force would result from an electron in the new cell being repelled by its own contribution to the charge in the old cell until the next update of Poisson's equation. The energy gain due to self-forces is exacerbated in the In$_{\text{0.53}}$Ga$_{\text{0.47}}$As $\Gamma$-valley as compared to Si. The ratio of energy gain between these two materials should roughly follow the ratio of the respective effective masses, which is in proportion to the product of the probability that a carrier will leave a grid site before Poisson's equation is updated, and how far it will likely travel in that remaining time period (where change in energy equals force times change in distance), both of which vary inversely with the square root of effective mass. Normally a self-force correction would be required to alleviate this artifact.~\cite{laux,aldegunde} However, with our use of sub-carriers, as detailed subsequently and whose impact is quantified below, the issue is nearly moot. The energy provided by the remaning effects of self-forces are small compared to the thermalizing effects on the carrier population by phonon scattering and the device contacts, as will be shown. 

The contacts are modeled by coupling the semiconductor S/D reservoirs to equilibrium electron distributions in the metal leads.  To realize perfectly injecting and absorbing contacts, we adjust the work function of the metal $q\Phi_{\mathrm{m}}$ to set the Fermi level $E_{\mathrm{F}}$ with respect to the conduction band edge $E_{\mathrm{C}}$ at the interface to provide a free electron carrier density corresponding to the doping density $N_{\mathrm{D}}$ within the S/D. Within the electron affinity rule, this means $q\Phi_{\mathrm{m}} = q\mathlarger{\mathlarger{\chi}} - (E_{\mathrm{F}} - E_{\mathrm{C}})$, where  $q\mathlarger{\mathlarger{\chi}}$ is the electron affinity of the semiconductor. This creates flat-band conditions at the contact interface. (The common value of $E_{\mathrm{F}} - E_{\mathrm{C}}$ is altered between simulations modeling classical and quantum statistics, however, for a given doping concentration.)  Warm-up periods of 2.4 picoseconds have proven sufficient to remove initial simulation transients. Final statistics were averaged over 18 picosecond intervals per gate bias.
\subsection{Silicon and III-V material models}
We generally follow the material parameters assembled in Refs.~\onlinecite{fischetti,jacoboni2} for valley-specific effective masses, non-parabolicity constants, and deformation potentials. We choose an analytic non-parabolic description of the band structure that accurately reproduces the DOS in Si up to 1.5 eV in the conduction band compared to full-band calculations,~\cite{rowlette} far larger than carrier energies produced by applied voltages of interest in our scaled devices. For Si, we model 6 ellipsoidal $\Delta$-valleys, while for In$_{\text{0.53}}$Ga$_{\text{0.47}}$As we include 1 $\Gamma$-, 4 L-, and 3 X-valleys. The $\Gamma$-valley is modeled as spherical while the L- and X-valleys are modeled as ellipsoidal.

We employ a virtual crystal approximation (VCA) to model $A_{\mathrm{x}}B_{\mathrm{1-x}}C$ ternary alloys such as In$_{\text{0.53}}$Ga$_{\text{0.47}}$As considered here. We specify bowing parameters for the intervalley separations between the $\Gamma$-, L-, and X-valleys, while all other parameters within the VCA are linearly interpolated. We typically model the intervalley separation $E_{\mathrm{\Gamma L}}$ between the light-mass $\Gamma$-valley and heavier-mass peripheral L-valleys as $E_{\mathrm{\Gamma L}} = 487$~meV, determined by a set of bowing parameters recommended by Vurgaftman and colleagues in their comprehensive review article.~\cite{vurgaftman} This value is a compromise between a commonly cited tight-binding calculation~\cite{porod} ($E_{\mathrm{\Gamma L}} = 460$~meV) and the only experimental determination~\cite{cheng} ($E_{\mathrm{\Gamma L}} = 550$~meV) to date. However, within the literature there is significant uncertainty in $E_{\mathrm{\Gamma L}}$.~\cite{regan} Recent density-functional calculations have estimated $E_{\mathrm{\Gamma L}}$ to be as large as 1.31~eV.~\cite{greene} Given such uncertainty, later we will analyze a fictitious In$_{\text{0.53}}$Ga$_{\text{0.47}}$As device having no satellite valleys whatsoever ($E_{\mathrm{\Gamma L}}\to\infty$). We will show that the main impacts of the peripheral valleys are to (i) increase the quantum capacitance via enhanced DOS, (ii) reduce the injection velocity due to heavier masses, and (iii) reduce the injection efficiency due to larger back-scattering. These effects are competing and it is not clear from the outset whether heavy occupation of the peripheral valleys will enhance or degrade device performance in III-V channels. 

For scattering, our simulator includes intravalley acoustic phonons within an elastic equipartition approximation and inelastic intra/intervalley deformation potential optical phonon scattering with a constant phonon energy.~\cite{jacoboni,jacoboni2} Umklapp $f$- and $g$-type intervalley scattering processes are included for Si~\cite{ferry4} and polar optical intravalley phonon scattering is considered for III-Vs.~\cite{jacoboni,jacoboni2} Degenerate ionized impurity scattering rates are calculated using a Brooks-Herring approach~\cite{brooks} employing a Thomas-Fermi screening model.~\cite{khess} We found this model to more readily reproduce low-field mobilities consistent with experiments in the degenerate limit compared to a Debye screening model. Alloy scattering is modeled with a crystal disorder deformation potential.~\cite{littlejohn} (SR scattering is included via QCs as discussed subsequently). We reproduced known bulk scattering rates as a function of energy for each scattering process individually to confirm our approach in each material.

We verified bulk transport by reproducing experimental carrier drift-velocity versus electric field curves for Si~\cite{jacoboni} and In$_{\text{0.53}}$Ga$_{\text{0.47}}$As~\cite{thobel,balynas} including the temperature dependence of the phonon bath at 300~K and 77~K. We not only matched the low-field mobilities but also the peak velocities in both materials to experimental data. Further, we verified that the onset of negative differential behavior in In$_{\text{0.53}}$Ga$_{\text{0.47}}$As, which denotes intervalley transfer of $\Gamma$-valley electrons to peripheral valley L-states, occurred at the correct electric field strength. Reproducing the velocity-field curves required small tunings of various deformation potentials, which is commonplace to MC simulation where deformation potentials are viewed as adjustable parameters.~\cite{fischetti} All our final simulation parameters and their references are listed in Appendix A.
\section{Quantum-corrections for non-equilibrium degenerate statistics}
Modern MOSFET devices employ large carrier concentrations throughout the device. With effective oxide thicknesses (EOTs) below 1 nm, and multi-gate geometries, large carrier concentrations can be obtained in the channel under gating in the ON-state. Activated S/D doping densities approaching solid-solubility limits then are used to improve performance by making the semiconductor reservoirs more metallic, reducing parasitic series S/D resistance, and increasing the overall device transconductance $g_{\mathrm{M}} = (\mathrm{d}I_{\mathrm{DS}}/\mathrm{d}V_{\mathrm{GS}})$. However, such carrier concentrations also can far exceed the conduction band effective DOS $N_{\mathrm{C}}$, raising the chemical potential well above the conduction band edge, the more so for lower DOS. Such degenerate carrier populations invalidate classical statistics (Boltzmann statistics in the equilibrium limit), and quantum statistics (Fermi statistics in the equilibrium limit) must be considered.

These quantum statistics are self-consistently produced by the consideration of what can simply be referred to as the PB of scattering. That is, the scattering rate $S(\mathbf{k}_i,\mathbf{k}_f)$ from any initial state $\mathbf{k}_i$ of occupation probability $f(\mathbf{k}_i)$ to some final state of occupation probability $f(\mathbf{k}_f)$ will be reduced in proportion to $1-f(\mathbf{k}_f)$ compared to what otherwise would be expected, 
\begin{equation}
S(\mathbf{k}_i, \mathbf{k}_f) = P(\mathbf{k}_i, \mathbf{k}_f)f(\mathbf{k}_i)\big(1 - f(\mathbf{k}_f) \big) ,
\label{equ:scat}
\end{equation}
to accommodate the Pauli exclusion principle. Here $P(\mathbf{k}_i, \mathbf{k}_f)$ is the scattering probability per unit time from a full state to an empty state. To address quantum statistics in otherwise semi-classical MC, with initial states intrinsically fully occupied in the MC method ($f(\mathbf{k}_i) = 1$), the PB of scattering typically is treated stochastically. Scattering events are first selected consistent with $P(\mathbf{k}_i, \mathbf{k}_f)$ pre-calculated by Fermi's Golden Rule. Then the scattering events are stochastically rejected with a probability $1-f(\mathbf{k}_f)$ according to the likelihood that the final state is already occupied. The question becomes what to use for the distribution function $f(\mathbf{k}_f)$, and how to determine it.

In a commonly employed approximation,~\cite{laux2,pantoja,mateos,islam,kalnaFD,islam2,david} the distribution function of final scattering states $f(\mathbf{k}_f)$ at position $\mathbf{r}$ is assumed to be a Fermi-Dirac distribution $f_{\mathrm{FD}}\big(\mathbf{r}, E(\mathbf{k}_f)\big)$ for the purposes of PB in Eq.~(\ref{equ:scat}). In this strategy, the shape of $f_{\mathrm{FD}}$ is determined by the local quasi-Fermi level and temperature consistent with the local carrier concentration and average energy. This approximation represents a great improvement over neglecting the PB of scattering when assuming classical statistics, being rigorous in the equilibrium limit. However, under strong non-equilibrium conditions including quasi-ballistic transport, actual distribution functions can become locally non-Fermi-like throughout the considered nanoscale device, including the channel. 

In this work, we make no \textit{a priori} assumptions about the shape of the distribution function. Instead, we calculate the distribution function to be used for PB directly by sampling the local carrier population $N(\mathbf{r}, E, g, \pm)$ as a function of position $\mathbf{r}$, valley $g$, energy $E$ relative to its respective valley-edge, and propagation directions forward toward the drain end ($+$) or backward toward the source end ($-$). The corresponding distribution function $f(\mathbf{r}, E, g, \pm)$ then is obtained from  
\begin{equation}
f(\mathbf{r}, g, E, \pm) = \frac{N(\mathbf{r}, g, E, \pm)}{D(g, E)/2} \;,
\end{equation}
where $D(g, E)/2$ is the position independent DOS per energy valley reduced by a factor of two for these half space ($\pm$) distributions. (In the case of quantum confinement considered below, $E$ is the energy referenced to, specifically, the non-quantum-corrected valley edge.) The resolution of $f(\mathbf{r}, E, g, \pm)$ in energy, $\Delta E$, is chosen depending on the equilibrium $E_{\mathrm{F}} - E_{\mathrm{C}}$ value in the S/D electron reservoirs. For light-mass In$_{\text{0.53}}$Ga$_{\text{0.47}}$As with an activated doping density of $N_{\mathrm{D}} = 5\times10^{19}~\mathrm{cm}^{-3}$ and a corresponding $E_{\mathrm{F}} - E_{\mathrm{C}} = 510$~meV, we choose an energy discretization of $\Delta E = k_{\mathrm{B}}T = 25.9$~meV at 300~K. For Si with a much greater DOS, at the same donor density we choose $\Delta E = k_{\mathrm{B}}T/4$ corresponding to a much smaller $E_{\mathrm{F}} - E_{\mathrm{C}} = 24.5$~meV. Si exhibits about $20\times$ greater quantum capacitance  $C_{\mathrm{Q}} = \mathrm{d}Q/\mathrm{d}(E_{\mathrm{F}}/q)$ than In$_{\text{0.53}}$Ga$_{\text{0.47}}$As when considering degenerate statistics.

Such approaches have been executed in $\mathbf{k}$-space for bulk calculations,~\cite{jacoboni,jacoboni2} but extension to device simulations has so far been prohibitive. This is because of large random-access memory demands and limited sample sizes.  We use three basic methods to increase our sample size: (i) averaging over short time periods, (ii) averaging over small regions in space and (iii) the use of fractional electrons or \textit{sub-carriers}. Considering (i), we average $N(\mathbf{r}, E, g, \pm)$ over a time period of 120~fs or 100~time steps, which is still an order of magnitude shorter that the switching time for even a THz transistor. With electrons moving only on the scale of Angstroms/time step, this time averaging effectively increases the sample size by roughly an order of magnitude.  For (ii), we average $N(\mathbf{r}, E, g, \pm)$ over the central nearest neighbor grid sites, increasing the sample size on average by a factor of 27 except at the device boundaries, at any given point in time. (Arguably, although not our motivation, averaging over nearest neighbor grid sites or beyond is perhaps more physically realistic than not doing so given the actual quantum-mechanical nature of the particles.) Finally considering (iii), we represent each real electron with 100 sub-carriers each carrying $1/100^{\mathrm{th}}$ the fundamental charge, increasing our sample size by another two orders of magnitude.  All told, the sample size from which we calculate $N(\mathbf{r}, E, g, \pm)$ and, thus, $f(\mathbf{r}, E, g, \pm)$, is effectively over four orders of magnitude larger than the physical number of carriers that would be expected at any grid point at any point in time!  

\begin{figure}[t]
	\includegraphics[width=0.75\columnwidth]{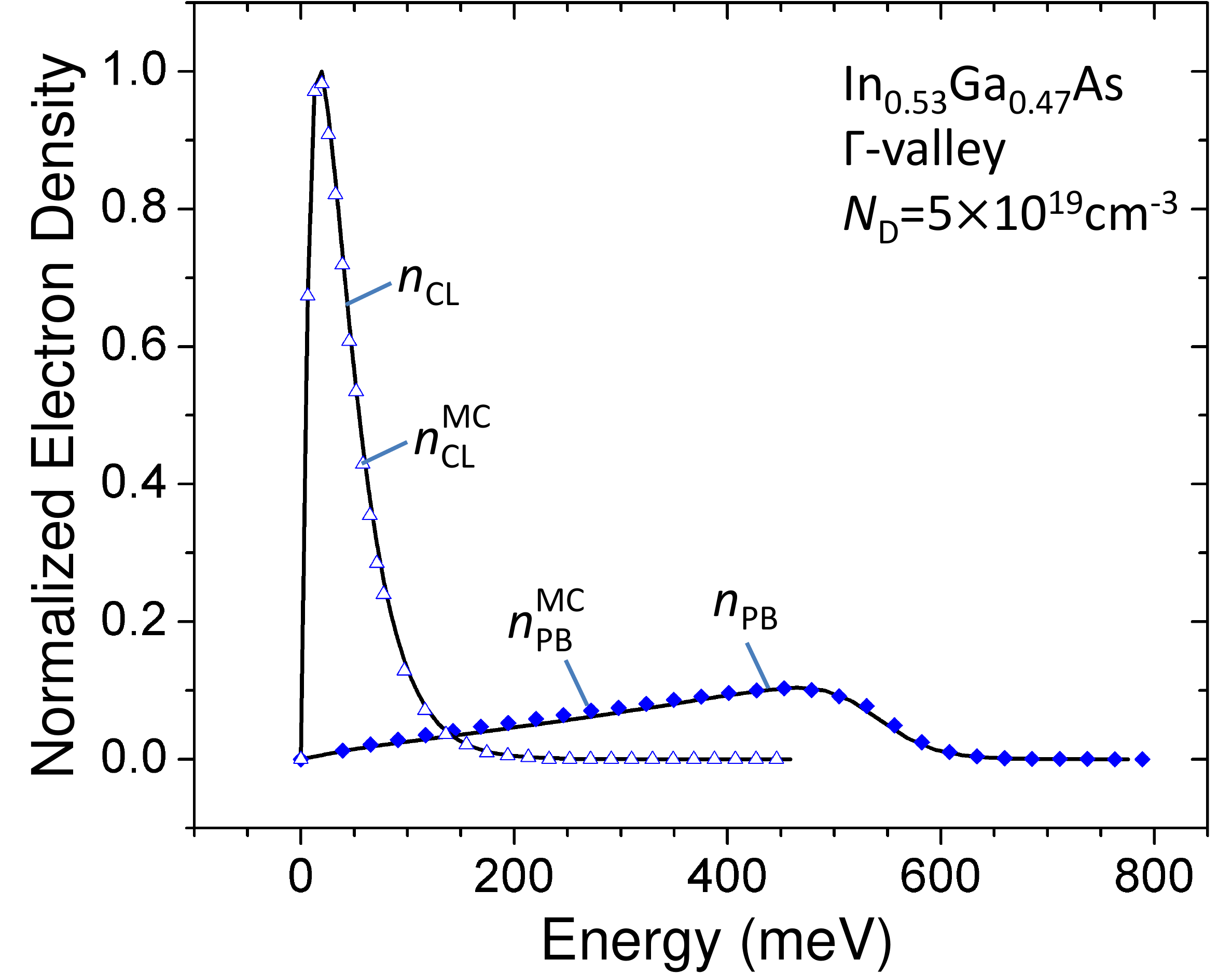}
	\caption{\label{fig:figPB} \textbf{\textbar\; Effects of degeneracy in In$_{\text{0.53}}$Ga$_{\text{0.47}}$As.} MC simulated electron distributions (symbols) with the PB of scattering ($n_{\mathrm{PB}}^{\mathrm{MC}}$) and without ($n_{\mathrm{CL}}^{\mathrm{MC}}$) in the $\Gamma$-valley of In$_{\text{0.53}}$Ga$_{\text{0.47}}$As. The results are shown under flat-band and equilibrium conditions, with reference distributions (solid lines) calculated directly from Fermi-Dirac ($n_{\mathrm{PB}}$) and Boltzmann ($n_{\mathrm{CL}}$) statistics, respectively. The distributions have been normalized to the peak electron density in the classical limit as a reference. There is $40\%$ less $\Gamma$-valley charge (area under the curve) in the degenerate case.}
\end{figure}

Fig.~\ref{fig:figPB} illustrates convergence of our PB method to the known equilibrium results in $\Gamma$-valley In$_{\text{0.53}}$Ga$_{\text{0.47}}$As electrons with a uniform device carrier concentration of $5\times10^{19}~\mathrm{cm}^{-3}$ at 300~K. (Since quantum effects are more pronounced in III-V materials, within this section and the upcoming Section III we illustrate our QCs in the In$_{\text{0.53}}$Ga$_{\text{0.47}}$As system, before returning to consider Si devices as well.) Here we closed the device boundaries and enforced flat-band conditions, considering only scattering and allowing the simulation to come to equilibrium. We then plotted the average local charge density distribution versus carrier kinetic energy, sampled over a single quasi-instantaneous 120~fs time interval. The carrier densities are normalized to the peak theoretical density in the classical limit. Our device populations both with ($n_{\mathrm{PB}}^{\mathrm{MC}}$) and without ($n_{\mathrm{CL}}^{\mathrm{MC}}$) the PB of scattering show excellent agreement compared to the reference equilibrium Fermi-Dirac ($n_{\mathrm{PB}}$) and Boltzmann ($n_{\mathrm{CL}}$)  statistics.  In In$_{\text{0.53}}$Ga$_{\text{0.47}}$As, the Fermi level rises nearly 400~meV into the conduction band upon considering degeneracy to accommodate the modeled $5\times10^{19}~\mathrm{cm}^{-3}$ carrier concentration.  This is in stark contrast to Si, where the Fermi energy only moves up 13~meV in the conduction band when considering PB at the given doping level. For In$_{\text{0.53}}$Ga$_{\text{0.47}}$As, the large change in the chemical potential occurs despite partial pinning of the Fermi level by the introduction of charge carriers into the peripheral L-valleys. L-valley electrons are not shown in Fig.~\ref{fig:figPB} for clarity. However, their occupation probabilities converge to the proper Fermi distribution as well, in concert with the $\Gamma$-valley. At the considered doping, the PB $\Gamma$-valley population is reduced to just $60\%$ of the total equilibrium density, with $40\%$ being in the L-valleys. 

While our sub-carriers significantly enhance our simulation statistics, our primary motivation for their use was to minimize classical carrier-carrier scattering which results from charge interacting via the time-dependent solution of Poisson's equation. This is in contrast to usual ensemble MC simulators, which typically embrace treating carrier-carrier scattering classically.  The ability to model carrier-carrier scattering via the Poisson solution is commonly thought to be a benefit of ensemble MC simulation versus full-quantum methods due to its simplicity and speed. However, classical molecular dynamics carrier-carrier scattering intrinsically neglects the PB of final-state pairs. Therefore, although these interactions serve to thermalize the carrier population, they do so towards a high-temperature Boltzmann distribution which is incompatible with non-equilibrium degenerate statistics. Moreover, the Coulomb force between two electrons at 2.5~nm apart (roughly the average separation for a carrier density of $5\times10^{19}~\mathrm{cm}^{-3}$) is quite strong at over 20~mV/nm, maximizing not only this classical thermalization effect, but also fictitious self-forces.  However, with our introduction of $N_{\mathrm{sub}}$ sub-carriers per real electron, each of which contributes only $q/N_{\mathrm{sub}}$ to the charge density, the Coulomb force among sub-carriers is reduced by $N_{\mathrm{sub}}$ (but not $N^2_{\mathrm{sub}}$) to 
\begin{equation}
\big|\mathbf{F}_{\mathrm{e-e}}\big| \simeq \frac{1}{N_{\mathrm{sub}}}\frac{q^2}{4\pi\varepsilon|\mathbf{r}_1-\mathbf{r}_2|} \;.
\end{equation}
This force still must be taken as proportional to the full charge on a real electron considering the local electric field to properly model the effects of the applied source, drain, and gate voltages.

Using the Golden Rule scattering rate as a measure, while the number of carriers to scatter off increases by $N_{\mathrm{sub}}$, the scattering rate between any two sub-carriers decreases by $N^2_{\mathrm{sub}}$, for a net reduction in the scattering rate by $N_{\mathrm{sub}}$. (Indeed, a future goal would be to introduce carrier-carrier scattering within a practical framework for which PB still can be considered.)  Thus we not only reduce the classical Coulomb force between sub-carriers but also their effective carrier-carrier scattering rates as well. 

\begin{figure}[t]
	\includegraphics[width=0.75\columnwidth]{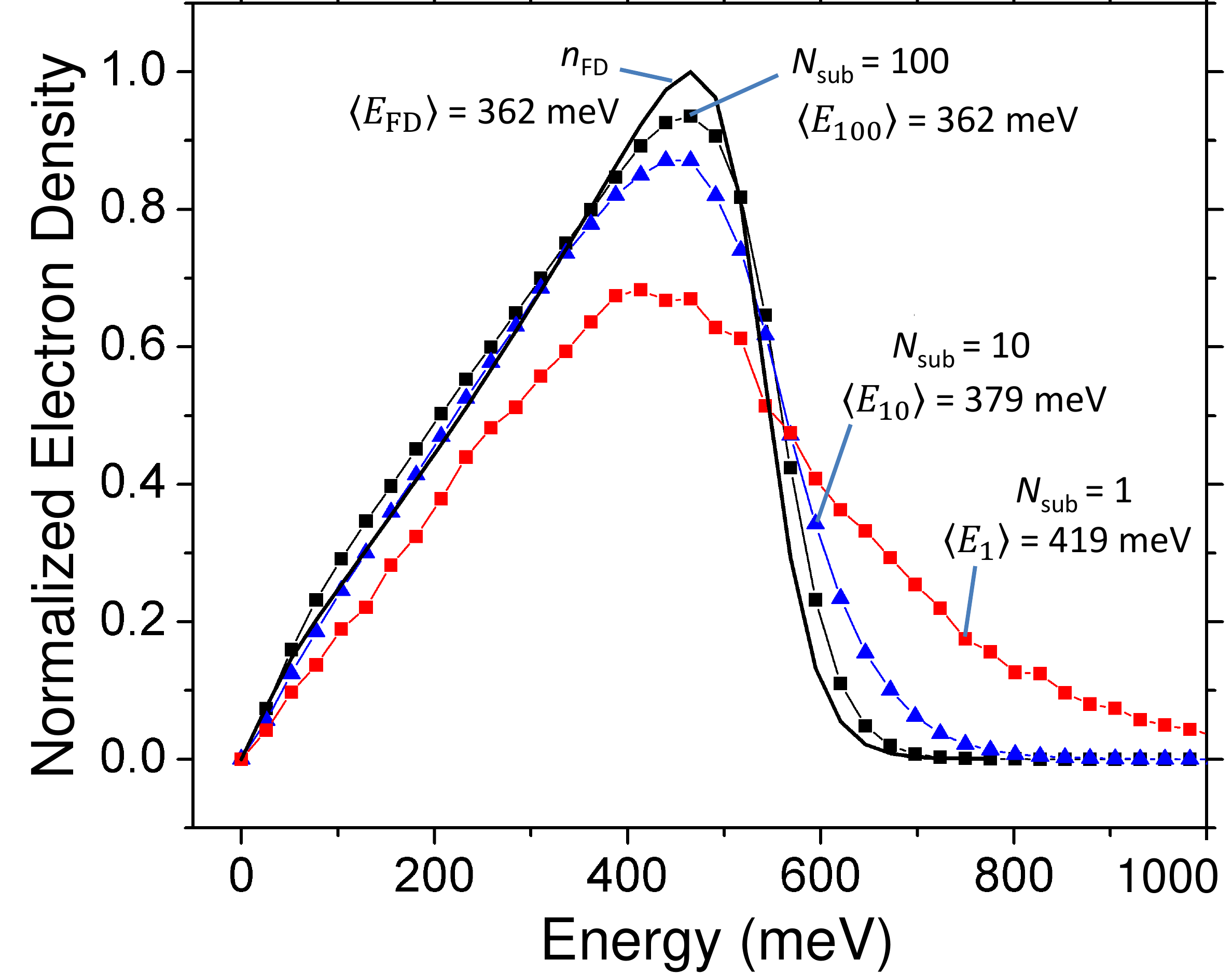}
	\caption{\label{fig:figHEAT} \textbf{\textbar\; Impact of sub-carriers on classical carrier-carrier interactions.} MC simulated electron distributions in the $\Gamma$-valley of In$_{\text{0.53}}$Ga$_{\text{0.47}}$As under equilibrium with different sub-carrier factors. The product of the 300~K Fermi-Dirac distribution and DOS is shown for reference $(n_{\mathrm{FD}})$. With $N_{\mathrm{sub}} = 1$, the unwanted thermalizing effects of classical carrier-carrier scattering drive the MC electron distribution to a high-temperature-like shape. Increasing the sub-carrier factor to $N_{\mathrm{sub}} = 10$ and then $N_{\mathrm{sub}} = 100$, the device distributions approach the theoretical expectation in shape and average kinetic energy $\langle E\rangle$. These distributions are normalized to have the same area under each curve. (The actual electron concentration in the $\Gamma$-valley is not conserved due to the occupation of peripheral valleys as a function of sub-carrier factor.)
}
\end{figure}
In Fig.~\ref{fig:figHEAT}, we turn on the self-consistent Poisson equation, open the S/D reservoirs to the metal contacts, and re-run the previous simulation study of Fig.~\ref{fig:figPB}, which was considered under flat-band conditions. The average local $\Gamma$-valley In$_{\text{0.53}}$Ga$_{\text{0.47}}$As charge density distribution versus carrier kinetic energy is shown, along with the average carrier kinetic energy per distribution, for differing sub-carrier factors $N_{\mathrm{sub}}$ to quantify their impact. With $N_{\mathrm{sub}} = 1$, classical carrier-carrier scattering thermalizes the electron population to an undesirable high-temperature-like distribution in energy, while self-forces raise the average kinetic energy about 60~meV above the theoretical expectation, despite the coupling to the boundaries and phonon scattering driving the electron distribution toward the Fermi-Dirac limit. However, as the sub-carrier factor is increased to 10, and then 100, the shape and average energy of the MC electron distributions approach those of the product of the DOS and the 300~K Fermi-Dirac distribution, $n_{\mathrm{FD}}$, and $\langle E_{\mathrm{FD}}\rangle$. This agreement is evidence that our sub-carrier strategy both mitigates the non-PB thermalization effects of classical carrier-carrier scattering, and relegates energy gains due to self-forces to negligible levels.  

\begin{figure}[t]
	\includegraphics[width=0.95\columnwidth]{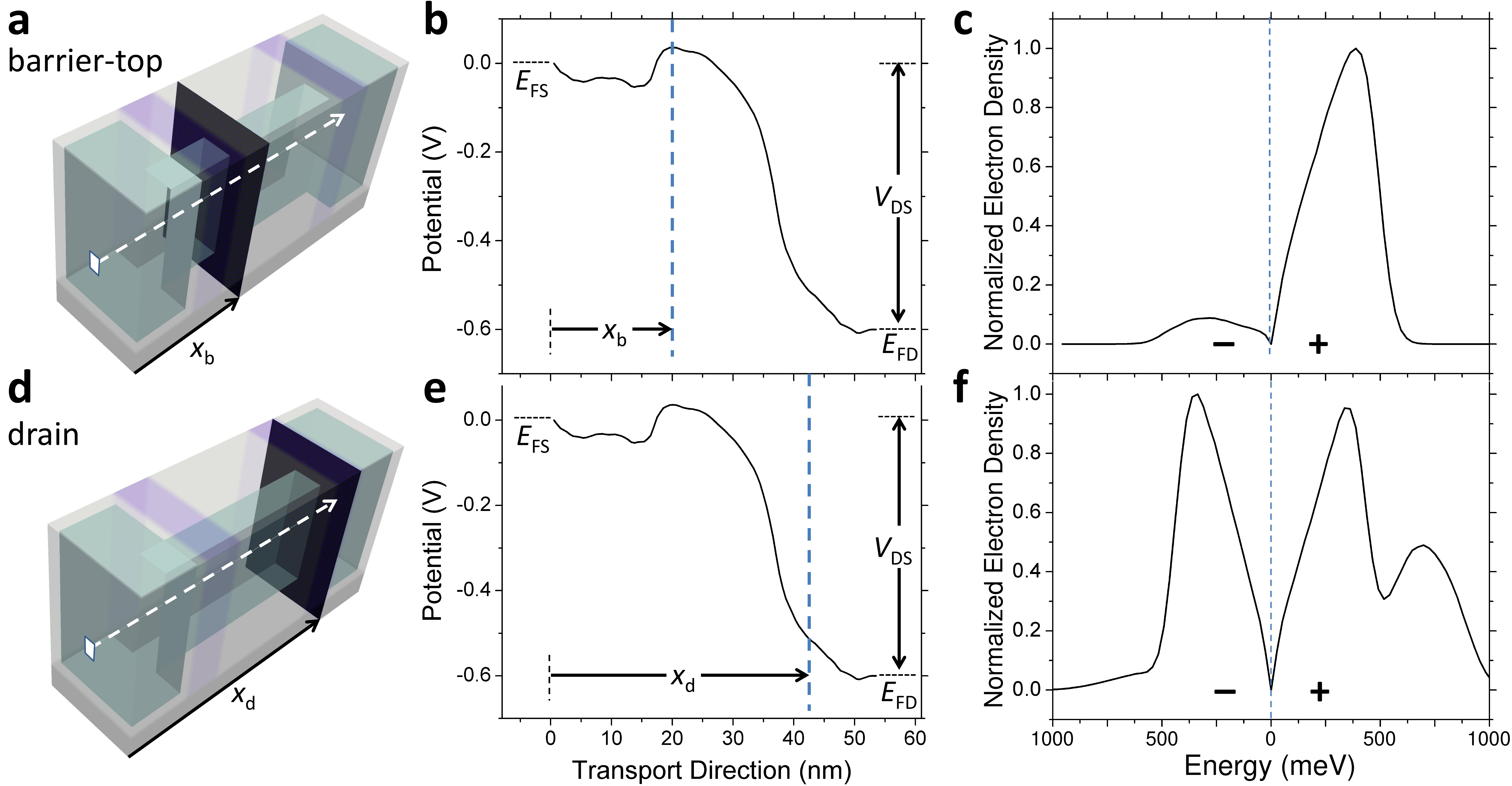}
	\caption{\label{fig:figocc} \textbf{\textbar\; Non-equilibrium electron distributions. (a)} FinFET edge view showing the device slice containing the potential energy barrier-top. \textbf{(b)} Potential profile along the dotted-white line of \textbf{a} showing the location of the barrier-top where the distributions of \textbf{c} were analyzed. \textbf{(c)} Normalized MC simulated electron distributions sampled at the potential energy barrier-top to the channel in the $\Gamma$-valley of In$_{\text{0.53}}$Ga$_{\text{0.47}}$As under non-equilibrium. The much larger forward-going contribution ($+$) versus backward-going ($-$) is consistent with a high injection efficiency into the channel. \textbf{(d)} FinFET edge view showing the device slice near the drain end of the channel. \textbf{(e)} Potential profile along the dotted-white line of \textbf{d} showing the location of the drain end where the distributions of \textbf{f} were analyzed. \textbf{(f)} Normalized MC simulated electron distributions again but sampled near the drain. The forward-going distribution ($+$) is the superposition of the drain near-equilibrium charge distribution and the quasi-ballistic population injected from the source.}
\end{figure}

The strength of our method, however, is that no \textit{a priori} assumption of an equilibrium\textemdash or any\textemdash specific distribution is made, in contrast to the use of a Fermi approximation for PB. We therefore conclude this section by illustrating our PB method under far-from-equilibrium conditions by sampling the In$_{\text{0.53}}$Ga$_{\text{0.47}}$As $\Gamma$-valley charge distributions under bias in Fig.~\ref{fig:figocc}. The bias conditions are source-to-drain voltage $V_{\mathrm{DS}} = 0.6$~V and gate overdrive above threshold $V_{\mathrm{ON}} = V_{\mathrm{GS}} - V_{\mathrm{T}} = 0.35$~V in accordance with ITRS predictions for future scaled MOSFETs.~\cite{*[{}] [{. Online: https://www.itrs.net.}] ITRS} The carrier distributions are sampled in the plane normal to the transport direction at the top of the channel potential-energy barrier-top (Fig.~\ref{fig:figocc}a) and also at the drain end of the channel (Fig.~\ref{fig:figocc}d). The electrostatic potentials, which are plotted along the dotted-white lines in Fig.~\ref{fig:figocc}a,d for reference, are visualized in Fig.~\ref{fig:figocc}b,e, where $x_{\mathrm{b}}$ represents the location of the barrier-top, and $x_{\mathrm{d}}$ the location of the beginning of the drain. The forward-going ($+$) and backward-going ($-$) carrier distributions necessarily differ greatly at the top of the channel barrier (Fig.~\ref{fig:figocc}c) consistent with a high injection efficiency. The forward-going distribution at the drain end (Fig.~\ref{fig:figocc}f) shows two peaks, the lower energy peak consistent with a nearly equilibrium distribution of charge carriers in the drain reservoir, and the higher energy peak consistent with quasi-ballistic electrons injected from the source.

It is clear from Fig.~\ref{fig:figocc} that the non-equilibrium electron occupation probability distribution is not consistent with any single Fermi distribution even as a local function of position. However, we cannot provide a direct comparison between the effects of the Fermi approximation versus our self-consistently obtained non-equilibrium distributions on device performance, such as on the drain current. The use of the Fermi approximation presents its own set of programming and computational challenges. Either sophisticated non-linear solvers must be used to determine the quasi-Fermi levels and temperatures, or large two-dimensional ($E_{\mathrm{F}}$, $T$) reverse look-up tables, e.g., ($n,\langle E\rangle)\to(E_{\mathrm{F}},T)$ would be required. Both strategies become increasingly difficult to use for highly degenerate statistics as considered in this work, where the average energy $\langle E\rangle$ becomes an increasingly weak function of $T$, and the numerical stability of the non-linear solvers becomes a concern. Thus, the inclusion of the Fermi approximation approach to degenerate statistics in our quantum-corrected MC simulator for the purpose of comparing methods is impractical.
\section{Quantum-corrections for electron quantum confinement}
Electrostatic quantum-correction potentials (QCPs) are widely used in MC simulation for the purpose of modeling quantum confinement. Each of our QCs for confinement described in this work are designed to employ the same set of QCPs. We employ first-principles QCPs which inherently require no fitting parameters as opposed to other methods like effective quantum potentials,~\cite{ferry,ferry2,ferry3} perturbative approaches,~\cite{heitz,heitz2} or density-gradient models~\cite{ancona,garcia} which require calibration. Specifically, we provide a valley-by-valley treatment of the space-, orientation-, and time-dependent QCPs based on the solutions of 2D effective mass Schr{\"o}dinger's equations solved in each transport slice.\cite{winstead,katha,fan,reg1,shi,shi2,david3,*[{}] [{ Ph.D. dissertation, The University of Texas at Austin.}] david2,lindberg,nagy} The valley and orientation dependence is provided by including the reciprocal effective mass tensor in the model Hamiltonian.~\cite{reg1,shi,shi2,david3,david2} This point is necessary to capture the self-consistent modification of intervalley separations and degeneracy splitting of otherwise equivalent valleys.

Our uses of the QCPs include altering energy separations between energy valley minima and calculating quantum-confinement-dependent phonon and surface-roughness scattering rates, in addition to redistributing charge carriers in real space and modifying source-to-channel potential barriers. (They are, however, not designed nor used to model quantum-mechanical tunneling-related leakage currents along the channel or through the gate.) Although it is our uses of the QCPs that are the focus of this work, we still describe their method of calculation here for clarity and completeness. 
\subsection{Obtaining the quantum-correction potential $\bm{V_{\mathrm{QC}}}$}
In general for this approach, for each valley $g$ (but not set of equivalent valleys) at position $\mathbf{r}$, the QCPs are defined by the relation 
\begin{equation}
\label{equ:rho}
\rho_{\mathrm{CL}}^{g} \big(\,\bar{V}(\mathbf{r})+V^g_{\mathrm{QC}}(\mathbf{r})\,\big) = \rho_{\mathrm{QC}}^{g} \big(\,\bar{V}(\mathbf{r})\,\big) \;.
\end{equation}
Here $V^g_{\mathrm{QC}}(\mathbf{r})$ is defined as the effective potential which, upon addition to the electrostatic potential $\bar{V}(\mathbf{r})$, will produce a classical device space-charge distribution $\rho_{\mathrm{CL}}$ equal to the quantum-mechanical one $\rho_{\mathrm{QM}}$. To smooth granularities in the potential found in the instantaneous device solutions, $\bar{V}(\mathbf{r})$ is a time-average over 100~time steps (120~fs total) of the potential $V(\mathbf{r})$ obtained from the self-consistent solution of Poisson's equation within the particle MC simulation, and includes the band and valley offsets in its definition. In equilibrium and for the 2D confinement considered here, within the $y-z$ plane of confinement, for each value of $x$ along the transport direction, Eq.~\ref{equ:rho} can be written as 
\begin{equation}
\label{equ:rho_2}
 \int\limits_{\bar{V}(\mathbf{r})+V^g_{\mathrm{QC}}(\mathbf{r})}^{+\infty}\!\!\!\!\!\!\!\!\!\!\mathrm{d}E\,D_{\mathrm{3D}}^g\big(E-\bar{V}(\mathbf{r})-V^g_{\mathrm{QC}}(\mathbf{r})\big)\,f_{\mathrm{FD}}(E) = \\
 \sum\limits_{i=1}^{M}\int\limits_{\bar{V}(\mathbf{r})}^{+\infty}\!\!\mathrm{d}E\,D_{\mathrm{1D}}^{g}\big(E-\bar{V}(\mathbf{r})\big)\,\big|\Psi^g_{i}(y,z)\big|^2\,\Big|_{x}f_{\mathrm{FD}}(E) \;.
\end{equation}
We include $M=20$~modes $i$ in the summation for the simulations of this work. $D_{\mathrm{1D}}^g$ and $D_{\mathrm{3D}}^g$  are the valley-wise 1D and 3D DOS, respectively. The $\Psi^g_{i}(y,z)$ are the 2D eigenvectors that diagonalize the Hamiltonian matrix $H$ within the 2D effective mass Schr{\"o}dinger equation, 
\begin{equation}
 H\,\Psi^g_{i}(y,z) = \big[\!-\frac{\hbar^2}{2}\,\nabla_{\perp}\cdot\frac{1}{\bm{m}^{\bm{*}}(y,z)}\cdot\nabla_{\perp}+\bar{V}(\mathbf{r})\big]\Psi^g_{i}(y,z) = E^g_i\,\Psi^g_{i}(y,z) \;,
\end{equation}
where 1/$\bm{m^*}$ is the reciprocal effective mass tensor, $\nabla_{\perp}=\hat{y}(\partial /\partial y) + \hat{z}(\partial /\partial z)$ is the transverse momentum operator, and $E^g_i$ are the valley- and sub-band-dependent eigenenergies. We solve the eigenvalue problem using a finite difference scheme that preserves the continuity of the probability current across the semiconductor interface. Here, the integrations of Eq.~\ref{equ:rho_2} are completed within the parabolic limit inside a Boltzmann approximation for the Fermi-Dirac distribution,~\cite{winstead,katha,fan,reg1,shi,shi2,david3,*[{}] [{ Ph.D. dissertation, The University of Texas at Austin.}] david2,lindberg,nagy} 
\begin{equation}
\label{equ:rho_3}
 N^g_{\mathrm{3D}}\,e^{\mathlarger{\frac{E_{\mathrm{F}}-\bar{V}(\mathbf{r})-V^g_{\mathrm{QC}}(\mathbf{r})}{k_{\mathrm{B}}T}}} =\\
 N^g_{\mathrm{1D}}\,e^{\mathlarger{\frac{E_{\mathrm{F}}-\bar{V}(\mathbf{r})-E^g_1}{k_{\mathrm{B}}T}}}\sum\limits_{i=2}^{M}\;\big|\Psi^g_{i}(y,z)\big|^2\,e^{\mathlarger{\frac{-E^g_i+E^g_1}{k_{\mathrm{B}}T}}} \;.
\end{equation}
Here, $N_{\mathrm{1D}}^g$ and $N_{\mathrm{3D}}^g$ are the valley-wise 1D and 3D effective DOS along the transport direction, respectively. The QCPs are revealed via logarithmic inversion as 
\begin{equation}
\label{equ:rho_4}
 V^g_{\mathrm{QC}}(\mathbf{r}) =\\
 E^g_1-k_{\mathrm{B}}T\,\mathrm{ln}\bigg(\frac{N^g_{\mathrm{1D}}}{N^g_{\mathrm{3D}}}\,\sum\limits_{i=2}^{M}\;\big|\Psi^g_{i}(y,z)\big|^2\,e^{\mathlarger{\frac{-E^g_i+E^g_1}{k_{\mathrm{B}}T}}}\bigg) \;.
\end{equation}
Finally, we then make a non-parabolicity correction to the QCPs by the reassignment
\begin{equation}
V^g_{\mathrm{QC}}(\mathbf{r}) \Leftarrow \frac{1}{\alpha_g}\bigg(\sqrt{\frac{1}{4}+\alpha_g V^g_{\mathrm{QC}}(\mathbf{r})}-\frac{1}{2}\bigg)\;,
\end{equation}
where $\alpha_g$ is the valley non-parabolicity constant. This correction is consistent with the reduction of the electron energy relative to the band edge due to non-parabolicity for a carrier of fixed wavelength, here defined by the quantum confinement.  Electrons, or sub-carriers here, are moved within the total potential according to the equations of motion 
\begin{equation}
\label{equ:force}
\mathbf{F}_g(\mathbf{r}) = \frac{d}{dt} \big( \hbar\mathbf{k} \big) = -q\nabla\cdot V_{\mathrm{tot}}(\mathbf{r}) = -q\nabla\cdot \big(V(\mathbf{r})+V^g_{\mathrm{QC}}(\mathbf{r}) \big),
\end{equation}
which govern the evolution of their crystal momentum. Consistent with our PB statistical updates and the time-averaging of $\bar{V}(\mathbf{r})$, our QCPs are updated every 120 fs in this work. 

There is no quantum-confinement in the S/D semiconductor regions since the electron wave functions can escape into the metal contacts. Thus there arises a question of how to approximate the 3D effects at the beginning and end of the conduction channel, where the quantum-confinement gradually turns-on and off moving from the unconfined S/D regions into and out of the restricted fin-channel. To estimate this transition, we first generated a $\Gamma$-valley QCP from a 2D slice \textit{along} the transport direction in the horizontal $x-y$ plane (not transverse). We observed an approximately linear turn-on in this QCP approaching the channel that was rougly equal to the physical width of the fin.  (The effective channel width allowing for barrier penetration is actually larger, which reduces the QCPs from what otherwise would be obtained.\cite{crum1})  To appoximate this effect, we linearly ramp-up the QCPs $V^g_{\mathrm{QC}}(\mathbf{r})$ from the outer edges of the drain and source extentions toward the gated channel over a distance $W_{\mathrm{fin}}$, the physical width of the channel in the horizontal plane.
\subsection{Uses of $\bm{V_{\mathrm{QC}}}$: real-space redistribution of charge}
We see from Eq.~\ref{equ:force} that the first effect of the QCPs is their application of classical forces on particles to redistribute them in real-space according to the as-calculated quantum-mechanical thermal charge distribution. This is provided to accurately model the capacitance of the gate, where it is known that the channel wave function is actually repelled from the oxide interface under quantum confinement.

\begin{figure}[t]
	\includegraphics[width=0.95\columnwidth]{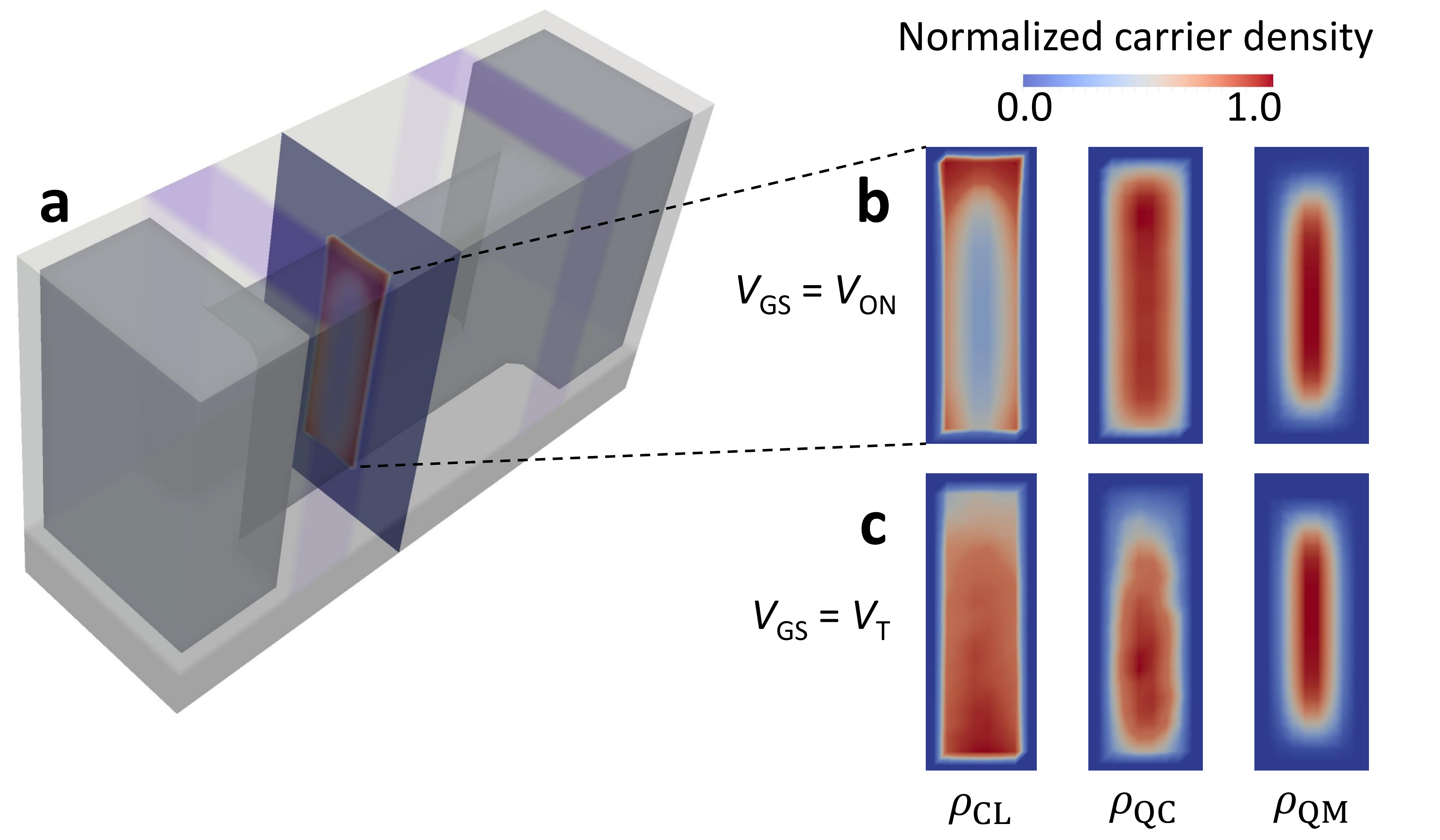}
	\caption{\label{fig:slice}\textbf{ \textbar\; Real-space redistribution of charge under quantum confinement. (a)} In$_{\text{0.53}}$Ga$_{\text{0.47}}$As FinFET with the channel slice containing the charge distributions of \textbf{b} and \textbf{c}. \textbf{(b)} At the overdrive gate voltage of 0.35~V above threshold, the classical electron charge density is attracted to the channel surface (top left, $\rho_{\mathrm{CL}}$). Including QCPs repels carriers from the interface (top middle, $\rho_{\mathrm{QC}}$) although degeneracy effects smear the spatial distribution relative to the as-calculated equilibrium Boltzmann-weighted quantum-mechanical charge density (top right, $\rho_{\mathrm{QM}}$). \textbf{(c)} Here $\rho_{\mathrm{QC}}$ (bottom middle) more closely resembles $\rho_{\mathrm{QM}}$ (bottom right) at threshold, but effects of degeneracy and non-equilibrium behavior are still evident. } 
\end{figure}

This spatial effect of the QCPs on the charge distribution is illustrated in Fig.~\ref{fig:slice} under 0.6~V drain bias for the channel cross-section of Fig.~\ref{fig:slice}a located near the beginning of the channel at the location of the potential energy barrier-top. We analyze the distribution at gate voltages of 0.35~V above threshold (Fig.~\ref{fig:slice}b) and at threshold (Fig.~\ref{fig:slice}c) in an In$_{\text{0.53}}$Ga$_{\text{0.47}}$As FinFET. Under each condition, we compare the purely classical MC device charge distribution ($\rho_{\mathrm{CL}}$), the MC device charge distribution including PB and quantum confinement ($\rho_{\mathrm{QC}}$), and the as-calculated equilibrium quantum-mechanical charge distribution in the Boltzmann limit from the preceding subsection ($\rho_{\mathrm{QM}}$, the right-hand side of Eq.~\ref{equ:rho_3}). 

Above threshold, a strong interface potential well attracts electrons to the surface, as expected classically ($\rho_{\mathrm{CL}}$, top left). Upon the inclusion of the QCPs, however, we see a device distribution with the population focused in the center of the channel and repelled from the interface as expected quantum-mechanically ($\rho_{\mathrm{QC}}$, top middle).  Yet this corrected shift of the carrier population is not as strong as for the reference Boltzmann equilibrium calculation ($\rho_{\mathrm{QM}}$, top right), nor should it be.  Unlike the calculations from which the QCPs were obtained, the quantum-corrected MC device simulations, corresponding to $\rho_{\mathrm{QC}}$, are also subject to degenerate statistics. Carrier degeneracy, combined with effectively reduced DOS for predominantly down-channel directed carriers (reduced by a factor of two in the ballistic limit), pushes carriers up in energy in the $\Gamma$-valley, as well as significantly into the peripheral valleys. This intervalley transfer of charge is due to a reduced $\Gamma$-L valley-splitting $E_{\mathrm{\Gamma L}}$. The satellite valleys have much larger effective masses, weaker quantum effects, and correspondingly smaller QCPs, as captured by our valley-by-valley treatment of the QCPs.  In In$_{\text{0.53}}$Ga$_{\text{0.47}}$As quantum wells, we typically see the quantum-corrected $E_{\mathrm{\Gamma L}}$ reduced by 200$-$300~meV in the channel of the considered FinFET depending on the voltage conditions. 

At threshold, but still under non-equilibrium degenerate conditions if less so, the quantum-corrected device distribution ($\rho_{\mathrm{QC}}$, bottom middle) looks more like the reference quantum-mechanical distribution ($\rho_{\mathrm{QM}}$, bottom right), but differences remain clear.  In this way, our use of the equilibrium Boltzmann approximation for the purposes of calculating the QCPs does not prevent us from using these same QCPs to address quantum confinement as applied to non-equilibrium degenerate carrier populations.   Moreover, our QCPs are most accurate but also most important for $E_{\mathrm{F}}$ near the quantum-corrected valley edges.  For example, the QCPs change the energy barrier heights for electrons to enter the constricted FinFET channel and, thus, in particular, the threshold condition. A good rule of thumb for estimating the resulting shift in threshold voltage $\Delta V_{\mathrm{T}}$ in the center of the channel due to quantum confinement is simply $\Delta V_{\mathrm{T}} \approx V_{\mathrm{QC}}^\Gamma$ for III-V materials (with a similar relation for Si considering $V_{\mathrm{QC}}^\Delta$).   For more energetic and more degenerate carrier populations, however, it is easier for electrons to reach the interface than would be expected in a fully quantum-mechanical calculation, so our QCPs remain somewhat conservative.
\subsection{Uses of $\bm{V_{\mathrm{QC}}}$: modeling confinement-dependent phonon and ionized-impurity scattering}
\begin{figure}[t]
	\includegraphics[width=0.95\columnwidth]{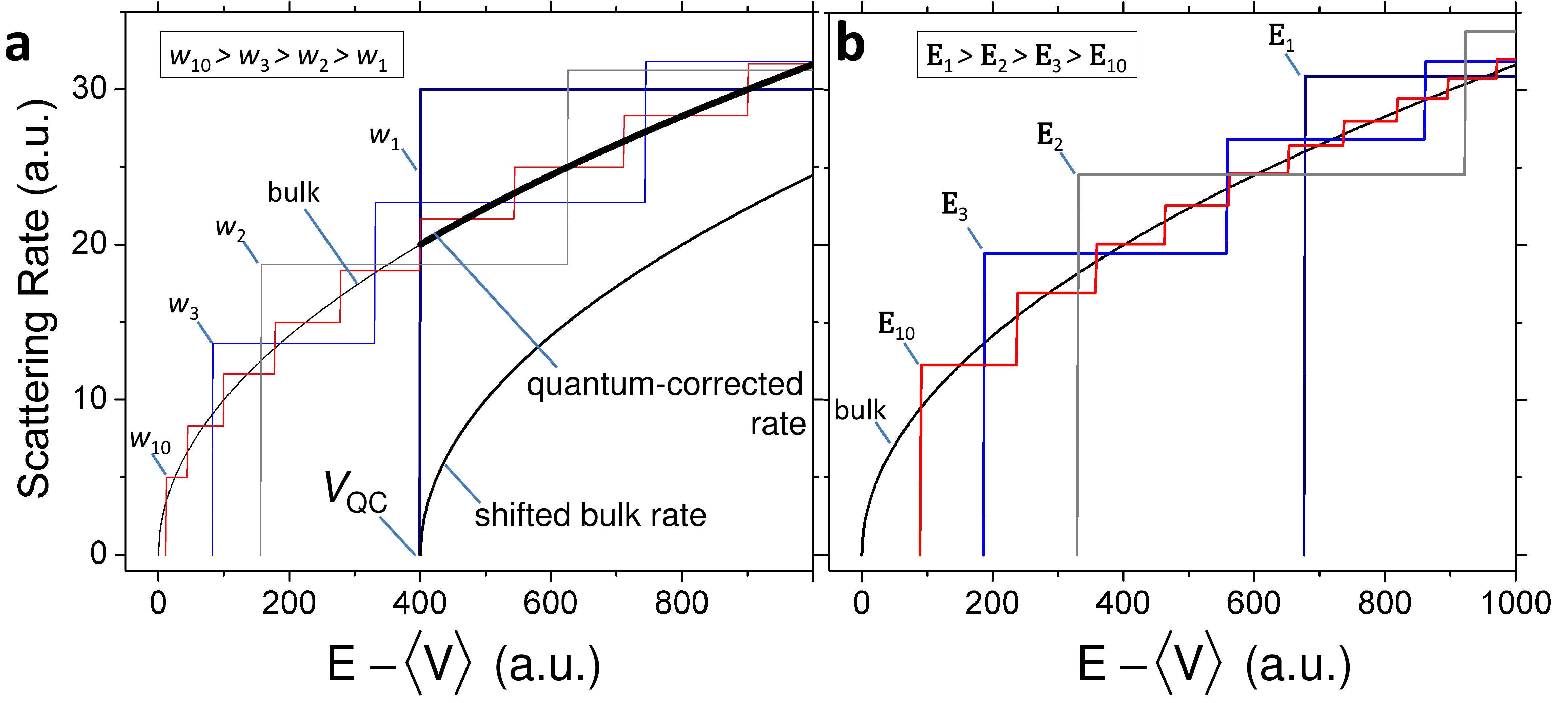}
	\caption{	\label{fig:osc}\textbf{ \textbar\; Modeling quantum-confined scattering using adjusted bulk rates.} The scattering rate for short range nominally randomizing elastic scattering within the first sub-band for 1D quantum confinement in \textbf{(a)} infinite square wells of varying width $w$, and \textbf{(b)} infinite one-sided triangular wells of varying fixed interface-normal field $\mathbf{E}$.  The curves are labeled by how many sub-bands are picked up within the given energy range for each respective confinement condition.  The scattering rates oscillate about the bulk rate (bulk) when the carrier energy is referenced to the expectation value of the electrostatic potential, more closely approaching the bulk limit with reduced confinement.  For increased confinement, the scattering rate for low-energy first sub-band carriers is increased substantially.  By using the bulk scattering rate but starting at the quantum-corrected band-edge $V_{\mathrm{QC}}$ (\textbf{a}, quantum-corrected rate), our approximation captures the overall increase in the basic scattering rate, if missing the oscillations and being somewhat conservative for low-energy carriers.  By comparison, simply shifting the zero of the bulk rate by the quantum-confined band-edge energy (\textbf{a}, shifted bulk rate) results in a much larger error. } 
\end{figure}

Strong quantum-confinement-enhanced phonon scattering long has been recognized, sufficient enough to more than halve the electron mobility in Si conduction channels with a few tenths of a MV/cm effective interface normal field absent even surface-roughness scattering.~\cite{yamakawa} This effect also has been seen in quantum transport calculations considering phonon scattering (as well as collision broadening thereof).~\cite{chen,kmliu} Scattering rates as a function of energy for electrons under quantum confinement oscillate about the bulk electron scattering rates with the introduction of each new final state sub-band when the confined carrier's energy is referenced to the expectation value of the uncorrected position-dependent valley edge. This behavior is illustrated via analytic Golden-Rule-based calculations for nominally randomizing short-range quasi-elastic (e.g., deformation potential acoustic phonon) scattering for one-dimensional confinement in an infinite square well and in a (one-sided) perfect triangular well in Fig.~\ref{fig:osc}a and b, respectively. (Initial and final state occupation probabilities have been neglected in these rates.) This illustrative calculation considers not only the modification of the final-state DOS but also the overlap between initial and final states which leads to a preference for intra-sub-band scattering while remaining otherwise randomizing. In the limit of wide wells and low interface fields, these results converge to the bulk limit as they must. For narrow wells and high interface fields, however, the scattering rate for the lowest energy carriers, which are substantially above the bottom of the square well or the expected potential for the triangular well due to quantum confinement, increases continuously as the confinement increases. 

To model these effects, we simply adjust the MC scattering rate by shifting the energy of the argument of the scattering rate for a given kinetic energy $\varepsilon_{\mathbf{k}}^i$ in the initial valley by the initial valley quantum correction $V_{\mathrm{QC}}^i$, as also illustrated (Fig.~\ref{fig:osc}a, quantum-corrected rate).  That is, the quantum-corrected scattering rate $R_{i\to f}^{\mathrm{\,QC}}$ from some initial state $i$ to some final state $f$ is given in terms of the uncorrected (classical) scattering rate $R_{i\to f}^{\mathrm{\,CL}}$ as
\begin{equation}
\begin{aligned}
\label{equ:scatRate}
& R_{i\to f}^{\mathrm{\,QC}}(\varepsilon_{\mathbf{k}}^i) = R_{i\to f}^{\mathrm{\,CL}}(\varepsilon_{\mathbf{k}}^i+V_{\mathrm{QC}}^i) & ;\;\;\; &\varepsilon_{\mathbf{k}}^f = (\varepsilon_{\mathbf{k}}^i +V_{\mathrm{QC}}^i - V_{\mathrm{QC}}^f - \Delta_{i,f} + \delta E )> 0 \;  \\
& R_{i\to f}^{\mathrm{\,QC}}(\varepsilon_{\mathbf{k}}^i) = 0  & ;\;\;\; &\varepsilon_{\mathbf{k}}^f = (\varepsilon_{\mathbf{k}}^i +V_{\mathrm{QC}}^i - V_{\mathrm{QC}}^f - \Delta_{i,f} + \delta E) < 0 \;
\end{aligned}
\end{equation}
for any assumed energy conserving scattering processes, where $\Delta_{i,f}$ is the uncorrected energy separation from the initial valley edge to the final valley edge, and $\delta E$ is the energy gained from (positive) or lost to (negative) the environment in the scattering process. We make the same adjustment for intravalley and intervalley scattering, quasi-elastic and inelastic, and (nominally) randomizing and non-randomizing scattering alike. (For intervalley scattering, the correction $V_{\mathrm{QC}}^i$ to the total energy of a carrier with kinetic energy $\varepsilon_{\mathbf{k}}^i$ raises the total energy with respect to the uncorrected band edge of the final valley just as for the initial valley.) Once the scattering rate has been selected, a specific final state subject to $\varepsilon_{\mathbf{k}}^f = \varepsilon_{\mathbf{k}}^i +V_{\mathrm{QC}}^i - V_{\mathrm{QC}}^f - \Delta_{i,f} + \delta E$ is chosen consistent with the relative probability determined by the difference in the initial and allowed final state $\mathbf{k}$ values in the usual way as appropriate for each scattering process. 

\begin{figure}[t]
	\includegraphics[width=0.75\columnwidth]{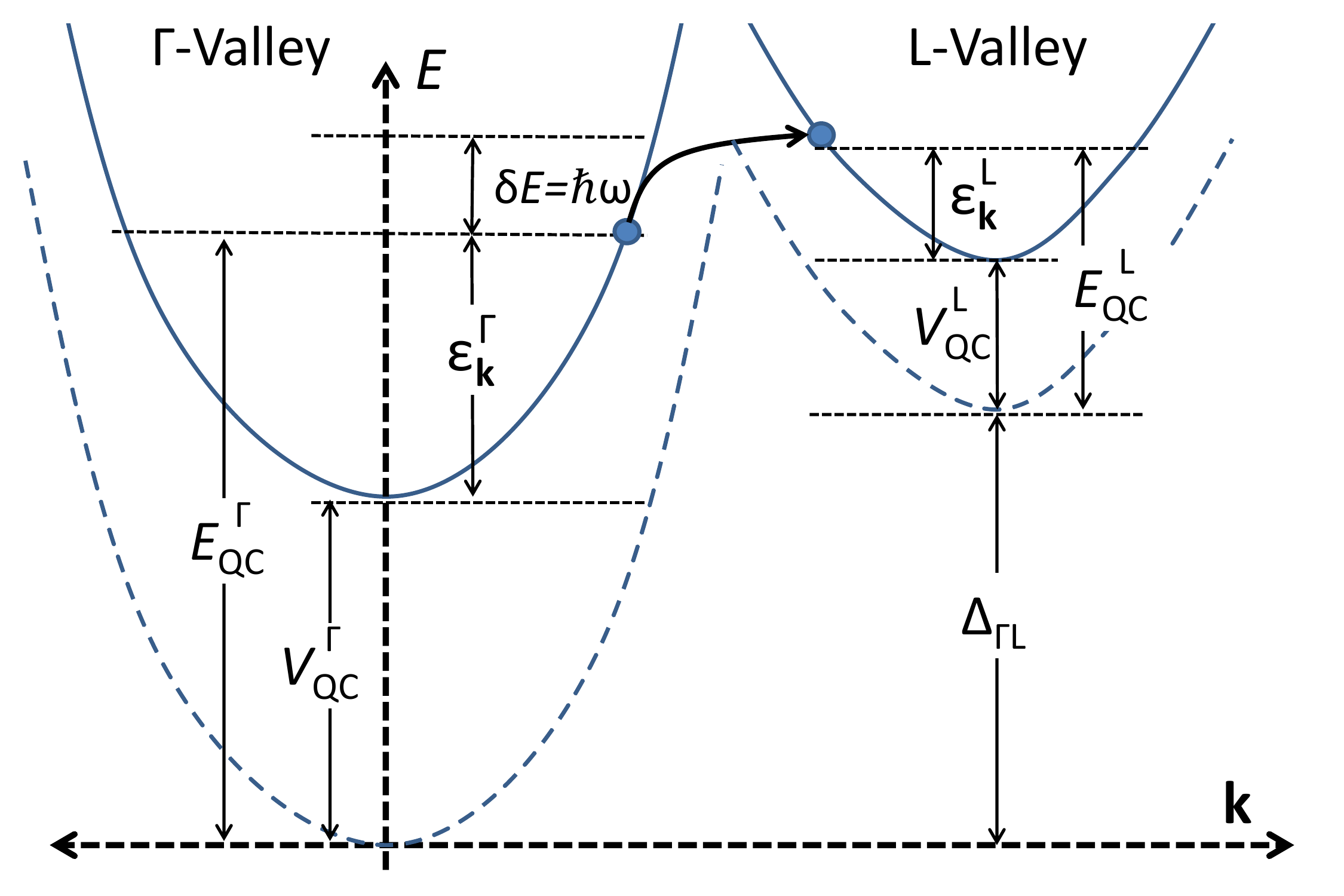}
	\caption{\label{fig:vPvqc}\textbf{ \textbar\; Selecting the choice of final state for quantum-confined scattering.} Quantum-confined intervalley scattering from the $\Gamma$-valley to an L-valley via absorption of an optical phonon of energy $\delta E = \hbar\omega$ in In$_{\text{0.53}}$Ga$_{\text{0.47}}$As, whereby the choice of final state scattering rate becomes a function of the total energy $E^{\Gamma}_{\mathrm{QC}} = \varepsilon_{\mathbf{k}}^{\Gamma} +V_{\mathrm{QC}}^{\Gamma}$, where the QCP $V_{\mathrm{QC}}^{\Gamma}$ raises the initial and, thus, final state energies relative to the uncorrected valley edges alike.} 
\end{figure}

While this approach misses the oscillations with energy and, in that way, is somewhat conservative for the lowest energy carriers, it captures the larger overall shift in the scattering rates with quantum confinement with no requirement to actually recalculate the scattering rates themselves from first-principles.  Certainly this approach is more effective than simply shifting the zero of the bulk scattering rate itself by the QCP (Fig.~\ref{fig:osc}a, shifted bulk rate). However we note that the effect of quantum confinement (and of this quantum correction in particular), can be relatively small or even reduce the scattering rates for non-randomizing long-range scattering interactions, consistent with their bulk energy dependence. An illustrative example of selecting the correct final quantum-corrected state energy for intervalley optical phonon absorption, and identifying the energies needed for the quantum-corrected scattering rate, is shown in Fig.~\ref{fig:vPvqc}. Consistent with the underlying localized-particle MC method, this correction is implemented in terms of the final-state valley QCPs as a local function of $\mathbf{r}$, $V_{\mathrm{QC}}^f(\mathbf{r})$. This approach has been used previously by our group~\cite{shi,shi2} for 1D confinement in 2D MC simulations, but it is extended here in this study to 2D confinement in 3D simulations.  

\begin{figure}[t]
	\includegraphics[width=0.75\columnwidth]{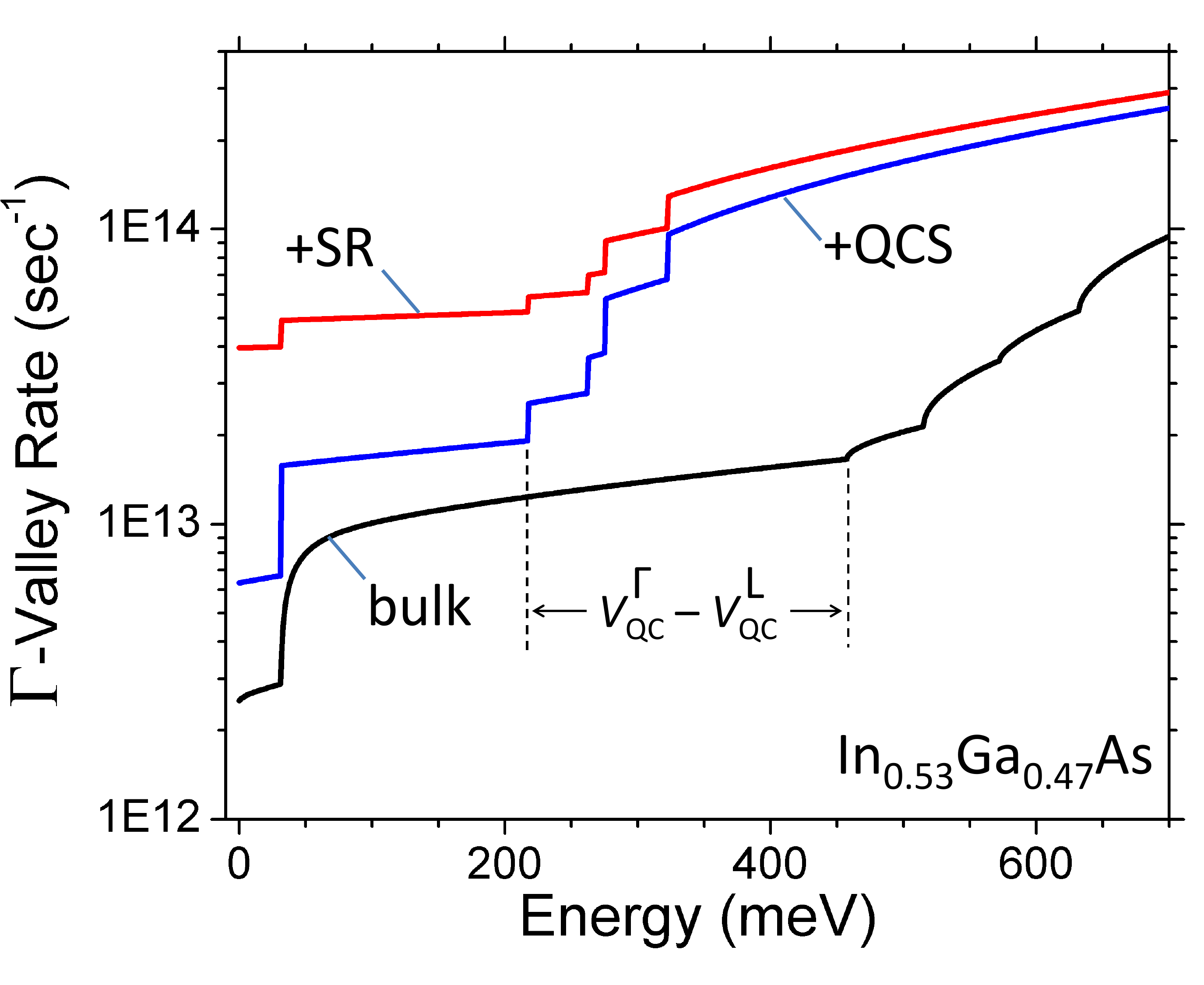}
	\caption{\label{fig:QCS}\textbf{ \textbar\; Enhanced scattering rates due to quantum confinement.} Calculated In$_{\text{0.53}}$Ga$_{\text{0.47}}$As $\Gamma$-valley scattering rates comparing the bulk rate against those including quantum-confined scattering (+QCS) and additional surface-roughness scattering (+SR) in the device channel center of a 6~nm wide fin under equilibrium conditions at threshold. Not only are the rates increased due to quantum confinement, but intervalley scattering, and thus the intervalley transfer (IVT) of electrons, occurs at lower kinetic energies versus the bulk case.} 
\end{figure}

In Fig.~\ref{fig:QCS} we illustrate our quantum-corrected scattering model applied to phonon scattering rates (+QCS), as well as the contribution from SR scattering (+SR) as discussed in the next section. The corrected rates are compared to the reference bulk rate (bulk) for electrons in the $\Gamma$-valley of In$_{\text{0.53}}$Ga$_{\text{0.47}}$As. The employed QCPs correspond to the channel center of a 6~nm wide fin geometry at equlibrium threshold conditions leading to moderately high values of $V_{\mathrm{QC}}^\Gamma = 375$~meV, and smaller values of $V_{\mathrm{QC}}^{\mathrm{L}}$ and $V_{\mathrm{QC}}^{\mathrm{X}}$, 150~meV and 100~meV, respectively. Not only are the rates of allowed scattering processes enhanced conistent with the $V_{\mathrm{QC}}^\Gamma$ shift in the energy argument, but the onset of intervalley scattering, and thus the intervalley transfer (IVT) of electrons, is reduced in energy by the difference in valley QCPs.   
\subsection{Uses of $\bm{V_{\mathrm{QC}}}$: modeling confinement-dependent surface-roughness scattering}
Surface-roughness (SR) scattering is calculated formally as a function of the effects of variation in the surface location and the quantum-confined energies, which makes it a candidate for approximation via QCPs. We motivate our final use of our position-, valley-, and orientation-dependent QCPs to model SR scattering by considering some limiting behaviors of SR scattering rates in well-known systems. In quantum wells defined by triangular confining potentials, such as in inversion layers in planar MOSFETs, the SR scattering rate varies as $S_{\mathrm{SR}}\propto E_{\mathrm{eff}}^2$ where $E_{\mathrm{eff}}$ is the interface-normal effective electric field defining the triangular well.~\cite{ando,ando2,ando3} Since the eigenenergies of the confined states $E_i$ have an $E_{\mathrm{eff}}^{3/2}$ dependence on the effective electric field, we note that the SR scattering rate varies as $E_i^{3}$.  In narrow infinite square wells of width $w$, SR scattering rates have been observed~\cite{jin,uchida} to obey $S_{\mathrm{SR}}\propto w^{-6}$, while the eigenenergies $E_i$ have a $w^{-2}$ dependence.  Again we note that SR scattering varies as $E_i^3$.  We therefore postulate, at least as a first \textit{ansatz}, a generalized approximate SR scattering rate for an electron at position $\mathbf{r}$ in valley $g$ as
\begin{equation}
\label{equ:SR}
S_{\mathrm{SR}}^g(\mathbf{r})\equiv \,C_{\mathrm{SR}}\times \big(V_{\mathrm{QC}}^g(\mathbf{r})\big)^3\;
\end{equation}
for our quantum-corrected MC simulations.  In addition to reproducing the basic confinement dependence of SR scattering in these two limits, this \textit{ansatz} also scales to the required result for very wide wells approaching the classical limit.  There the SR is only significant for carriers near the surface and where our QCPs also would remain nonzero, modeling quantum-mechanical surface repulsion which remains no matter how wide the well, but is otherwise width independent.  

In Eq.~\ref{equ:SR}, while $\big(V_{\mathrm{QC}}^g(\mathbf{r})\big)^3$ represents the effects of quantum confinement on SR scattering, the lead coefficient $C_{\mathrm{SR}}$ represents the amount of actual SR.  SR is a function of interface quality and, thus, will vary with both channel and dielectric material and even strain and growth conditions, particularly in nascent technologies. However, the lack of universal experimental results and rapid evolution in both FinFET and III-V technologies makes a calibration of $C_{\mathrm{SR}}$ for these systems problematic. For this work, we calibrated $C_{\mathrm{SR}}$ to reproduce known experimental SR scattering rates obtained for a planar Si/thermal SiO$_2$ interface channel, having also considered confined phonon scattering in both cases.~\cite{yamakawa}  This approach likely is optimistic for Si FinFETs and, more so, for In$_{\text{0.53}}$Ga$_{\text{0.47}}$As FinFETs.  However, it also provides another control for our simulations, allowing us to focus on the intrinsic properties of the two host material systems.  However, although the assumed SR represented by $C_{\mathrm{SR}}$ is held constant, the behavior of the quantum confinement\textemdash represented here via the QCPs\textemdash and, thus, the actual SR \textit{scattering} will vary with material as well as energy valley and valley orientation.  

Here, since the interface roughness interaction with the channel wave function is non-local along the channel, we have chosen an elastic long-range non-randomizing polar optical phonon-like selection procedure for the final state after SR scattering,~\cite{jacoboni,jacoboni2} calibrated as noted above, which should be sufficient for the purposes of this work. We have plotted the thus-calculated SR scattering rate contribution for the electrons in the $\Gamma$-valley of In$_{\text{0.53}}$Ga$_{\text{0.47}}$As of the same FinFET channel in Fig.~\ref{fig:QCS}.  Initially the SR scattering more than doubles the scattering rate despite the likely underestimated amount of SR. However, with increasing energy, the randomizing intervalley phonon scattering processes soon become the dominant scattering mechanisms again.
\section{Model comparison and discussion}
\begin{figure*}[t!]
	\includegraphics[width=0.95\textwidth]{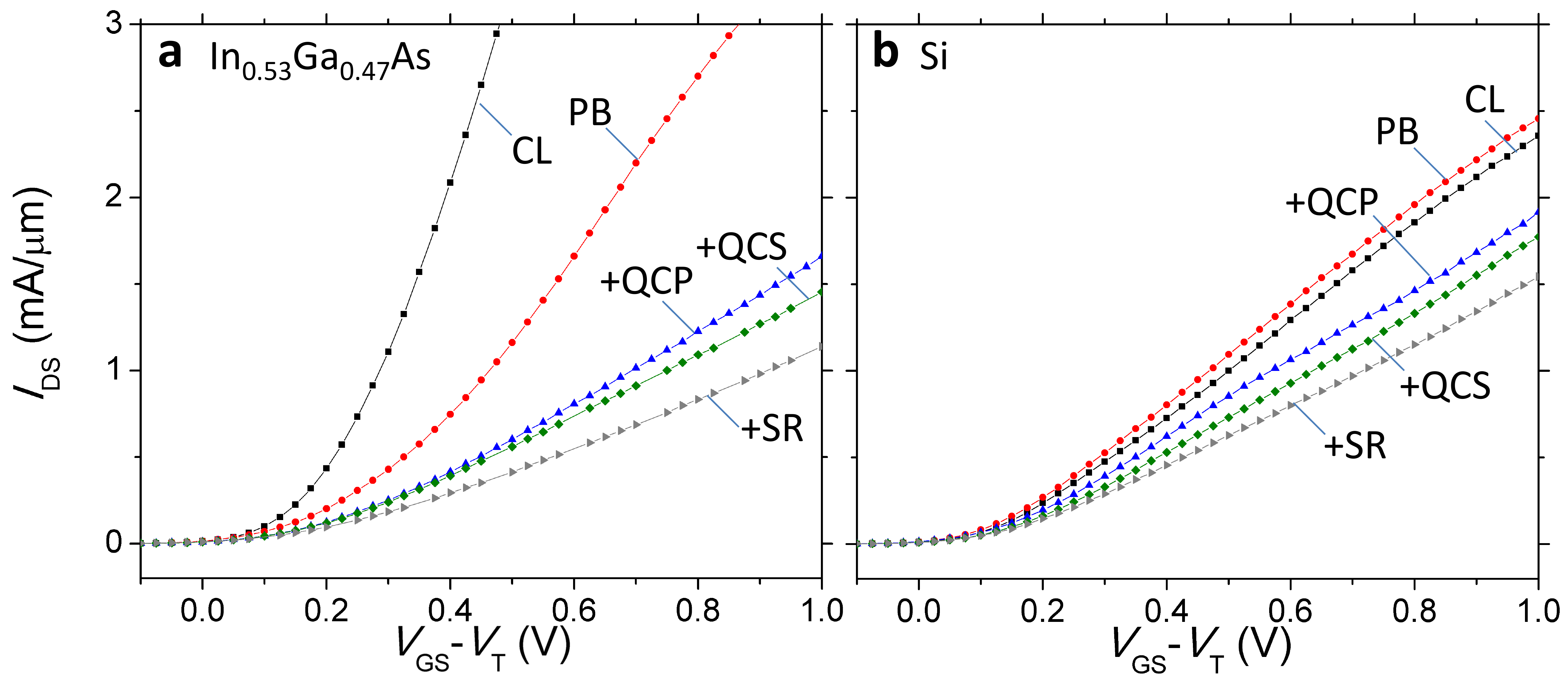}
	\caption{\label{fig:model}\textbf{ \textbar\; $\mathbf{(100)/\langle100\rangle}$ FinFET performance comparison of In$_{\text{0.53}}$Ga$_{\text{0.47}}$As versus Si with varying quantum-corrected models. (a)} As the Pauli-blocking (PB) of scattering, and then quantum-correction potentials (+QCP) are added, as well as QC-dependent phonon (+QCS) and surface-roughness scattering (+SR), the performance of the In$_{\text{0.53}}$Ga$_{\text{0.47}}$As FinFET device (left) is substantially degraded relative to classical (CL) expectations. \textbf{(b)} Si is shown to be more robust against carrier degeneracy and quantum-confinement effects (right), although its performance is also moderated compared to classical expectations.} 
\end{figure*}

We illustrate the QCs, and the effects of degenerate populations and quantum confinement, by benchmarking In$_{\text{0.53}}$Ga$_{\text{0.47}}$As devices against industry standard Si devices, and by isolating the impact of each quantum correction. The FinFET sidewall orientation is $(100)/\langle 100\rangle$ and the reference FinFET geometry is described in detail in Section II. We compare the two materials' respective transfer curves $I_{\mathrm{DS}} - V_{\mathrm{GS}}$ in Fig.~\ref{fig:model} and their transconductances and ON-state currents in Table I sampled at the overdrive gate voltage of $V_{\mathrm{ON}} = V_{\mathrm{GS}} - V_{\mathrm{T}} = 350$~mV, while adding modeled quantum effects one at a time. We also analyze these results in terms of a quasi-ballistic representation of the current motivated by Lundstrom~\cite{L1,L2}
\begin{equation}
I_{\mathrm{DS}} = q\,n_{\mathrm{b}}\,v_{\mathrm{inj}} \,\gamma \;,
\end{equation}
where $q$ is the fundamental charge and, by definition, $n_{\mathrm{b}}$ is the total cross-sectional charge density at the top of the source-to-channel potential barrier. The injection velocity $v_{\mathrm{inj}}$ is the average velocity along the channel of incident charge carriers (those moving toward the drain) at the barrier-top, and $\gamma$ is the injection efficiency. The injection efficiency is    
\begin{equation}
\gamma = \frac{1-R_j}{1+R_n} \;,
\end{equation}
with a distinction made here between reflection in terms of current $j$ reflection in the numerator and charge $n$ reflection in the denominator. The ratio of reflected ($-$) current to incident ($+$) current is $R_j \equiv \langle n_-\rangle \langle v_-\rangle\mathlarger{\mathlarger{/}} \langle n_+\rangle\langle v_+\rangle$, the ratio of reflected charge to incident charge is $R_n = \langle n_-\rangle\mathlarger{\mathlarger{/}} \langle n_+\rangle$, $\langle v_+\rangle \equiv v_{\mathrm{inj}}$, and $\langle n_+\rangle + \langle n_-\rangle  \equiv n_{\mathrm{b}}$, with all parameters once again measured at the barrier-top. Meanwhile, $n_{\mathrm{b}}$ depends on the gate overdrive voltage, controlled by the series combination of the dielectric capacitance and the channel quantum (DOS) capacitance.   

\begin{table}[t!]
\begin{center}
	\begin{tabular}{ | l c c c c c c |}  
	\multicolumn{7}{c}{\textbf{\textlarger{ In$_{\text{0.53}}$Ga$_{\text{0.47}}$As}}}		\\						
	\hline
	&\quad\;					&	\quad CL \quad	&	\quad PB \quad		&	\quad +QCP \quad 		&	\quad +QCS \quad			&	\quad +SR\quad	\quad		\\
	\hline
	&\quad$g_{\mathrm{M}}$ (mA/$\mu$m/V)		&	\quad9.90\quad	&	\quad3.17\quad		&	\quad1.62\quad			&	\quad1.50\quad				&	\quad1.09\quad	\quad		\\

	&\quad$I_{\mathrm{ON}}$ (mA/$\mu$m)		&	\quad1.57\quad	&	\quad0.57\quad		&	\quad0.33\quad			&	\quad0.31\quad				&	\quad0.23\quad	\quad		\\
	\hline

	\multicolumn{7}{c}{\textbf{\textlarger{Si}}}		\\						
	\hline
	&\quad			&	\quad CL \quad	&	\quad PB \quad		&	\quad +QCP \quad 		&	\quad +QCS \quad			&	\quad +SR\quad	\quad		\\
	\hline
	&\quad$g_{\mathrm{M}}$ (mA/$\mu$m/V)		&	\quad2.51\quad	&	\quad2.71\quad		&	\quad2.30\quad			&	\quad2.04\quad				&	\quad1.63\quad	\quad		\\

	&\quad$I_{\mathrm{ON}}$	(mA/$\mu$m)	&	\quad0.60\quad	&	\quad0.67\quad		&	\quad0.50\quad			&	\quad0.43\quad				&	\quad0.37\quad\quad			\\
	\hline

\end{tabular}
\caption{\textbf{ \textbar\;$\mathbf{ON}$-state performance by model.} The transconductance $g_{\mathrm{M}}$ and the drain current $I_{\mathrm{ON}}$, sampled at $V_{\mathrm{DS}} = 600$~mV and $V_{\mathrm{ON}} = V_{\mathrm{GS}} - V_{\mathrm{T}} = 350$~mV. Both have been normalized by the fin perimeter, $(2T_{\mathrm{fin}}+W_{\mathrm{fin}}$). } 
\end{center}
\end{table}
In$_{\text{0.53}}$Ga$_{\text{0.47}}$As strongly outperforms Si under classical MC simulation assumptions (Fig.~\ref{fig:model}a and b, CL), consistent with a smaller transport effective mass in its $\Gamma$-valley than in the six equivalent Si $\Delta$-valley carriers. Since degeneracy is not considered in our CL model, both devices exhibit similar gate capacitances, and the 3$\times$ difference in $I_{\mathrm{ON}}$ can be explained by the roughly 3$\times$ larger injection velocity of In$_{\text{0.53}}$Ga$_{\text{0.47}}$As versus Si in the semi-classical  limit. The observed injection velocity of In$_{\text{0.53}}$Ga$_{\text{0.47}}$As is curbed, however, by the large non-parabolicity constant $\alpha_{\Gamma}$ of its $\Gamma$-band, especially at larger kinetic energies, where $\alpha_\gamma$ can decrease incident velocities as much as 50$\%$.

The inclusion of the PB of scattering and resulting degenerate statistics then greatly moderates the performance of In$_{\text{0.53}}$Ga$_{\text{0.47}}$As relative to Si (Fig.~\ref{fig:model}a and b, PB). Indeed, PB-Si actually performs better than in the CL-Si case due to increased thermal velocities of degenerate electrons injected from the source versus the non-degenerate CL case. Although the same is true of In$_{\text{0.53}}$Ga$_{\text{0.47}}$As, whose injection velocity increases from 3.2 to $4.3\times10^7$~cm/sec considering degeneracy, its cross-sectional charge density is reduced 85$\%$ in the PB case due to severely reduced quantum (DOS) capacitance in this light-effective mass material. 

In addition to degeneracy, we then employ the quantum-correction potentials (Fig.~\ref{fig:model}a and b, +QCP) to model our first level of quantum confinement. The performance of In$_{\text{0.53}}$Ga$_{\text{0.47}}$As is diminished as the charge density is redistributed among the energy valleys through scattering via reduced intervalley separations. In the PB case, $99\%$ of the sampled carriers at the barrier-top occupy $\Gamma$-states, which is reduced to only $26\%$ in the +QCP case. This corresponds to a reduction in the injection efficiency from $\gamma_{\mathrm{PB}}=90\%$ down to $\gamma_{\mathrm{QCP}}=43\%$, coinciding with the greater occupation of heavier-mass satellite valleys, which experience worse backscattering. Notably the thermal velocity also dwindles from 4.3 down to $2.5\times10^7$~cm/sec.  However, there is a competition here, as occupation of the L- and X-valleys rapidly increases the differential quantum capacitance of In$_{\text{0.53}}$Ga$_{\text{0.47}}$As, and gate control over $n_{\mathrm{b}}$, correspondingly. Si is more robust to quantum confinement in contrast, however. In this $(100)/\langle100\rangle$ fin configuration, the confinement-induced degeneracy-splitting of the $\Delta$-bands results in larger occupation of the $\Delta_2$-valleys oriented normal to the sides of the fin, which have the lightest transport effective mass and higher thermal velocity. This partially offsets worse quantum (DOS) capacitance, leading to an overall smaller relative reduction in current for Si compared to In$_{\text{0.53}}$Ga$_{\text{0.47}}$As. These results in both materials highlight the importance of a valley-by-valley treatment of quantum corrections.

We next add the quantum-confined scattering using the same QCPs (Fig.~\ref{fig:model}a and b, +QCS). Performance degrades in both material systems due to significantly enhanced scattering rates, although this quantum correction had the smallest effect on the overall drive current performance in the considered device structure.

Finally, we add the surface-roughness scattering (Fig.~\ref{fig:model}a and b, +SR). $I_{\mathrm{ON}}$ and $g_{\mathrm{M}}$ suffer in both materials, albeit worse for In$_{\text{0.53}}$Ga$_{\text{0.47}}$As. The light-mass $\Gamma$-valley carriers, having the largest QCPs, experience the worst SR scattering, reducing the In$_{\text{0.53}}$Ga$_{\text{0.47}}$As overall injection efficiency to $\gamma_{\mathrm{SR}}=33\%$ down from $\gamma_{\mathrm{QCS}}=43\%$.  Nevertheless, the effect of SR scattering is smaller than one might otherwise expect, even allowing for the proximity to the ballistic limit. Previously our group has observed in both MC~\cite{shi,shi2} and quantum transport simulations~\cite{kmliu} that SR and phonon scattering rates are not simply additive concerning the effects on channel transport, and that the whole is less than the sum of the parts. (E.g., phonon emission reducing a carrier's energy to below the barrier-top can prevent subsequent SR induced back-scattering.) Also, because of calibration to the amount of SR to the more ideal planar Si-SiO$_2$ interface, the estimates of SR and, thus, SR scattering, therefore, are likely conservative.

In terms of the effects of occupation of the peripheral valleys in In$_{\text{0.53}}$Ga$_{\text{0.47}}$As, there is ambiguity for two reasons, however.  First, at the considered valley offset of $E_{\mathrm{\Gamma L}} = 487$~meV and doping density of $N_{\mathrm{D}} = 5\times10^{19}~\mathrm{cm}^{-3}$, the Fermi level is sufficiently high in the source and drain that we inject directly into the bottom of the satellite L-valleys in the simulations above. Thus the mechanism by which carriers reach the peripheral valleys within the channel, whether by this injection from the boundaries or by scattering within the simulation region, is inconclusive. Therefore the role of intervalley scattering on transport is unclear. Moreover, there is significant uncertainty in reported intervalley separations, with some models suggesting~\cite{greene} sufficiently large separations that carriers would remain almost entirely localized to the $\Gamma$-valley.  We address both of these issues by reducing the assumed source and drain doping to $1\times10^{19}~\mathrm{cm}^{-3}$, below that achievable through post-crystal growth doping, such that the Fermi level at $E_{\mathrm{F}} - E_{\mathrm{C}} = 290$~meV remains well below the peripheral valley edges and eliminates injection into the L-valleys from the contacts. We then perform simulations with and without intervalley scattering to isolate its impact. Including intervalley scattering provides understanding of the role of intervalley scattering within the simulation region on transport provided the peripheral valleys are available for occupation. Excluding intervalley scattering allows analysis of device behavior absent the availability of peripheral valleys for occupation. 

With the intervalley separations as considered previously, intervalley scattering turned on, but the S/D doping reduced to $1\times10^{19}~\mathrm{cm}^{-3}$, we still find large occupation of the satellite valleys in the channel in the ON-state, where the gate and not the S/D doping controls the carrier density and the intervalley separation is reduced by quantum confinement. Roughly $58\%$ of the charge in the channel occupies L- and X-states, although all the charge from the source and drain is injected into the $\Gamma$-valley. This result points to the continuing need to model intervalley scattering within the simulation region\textemdash absent contact injection, and perhaps even with it, ballistic treatments of transport cannot model the role of energetically available peripheral valleys. 

\begin{figure*}[t!]
	\includegraphics[width=0.95\textwidth]{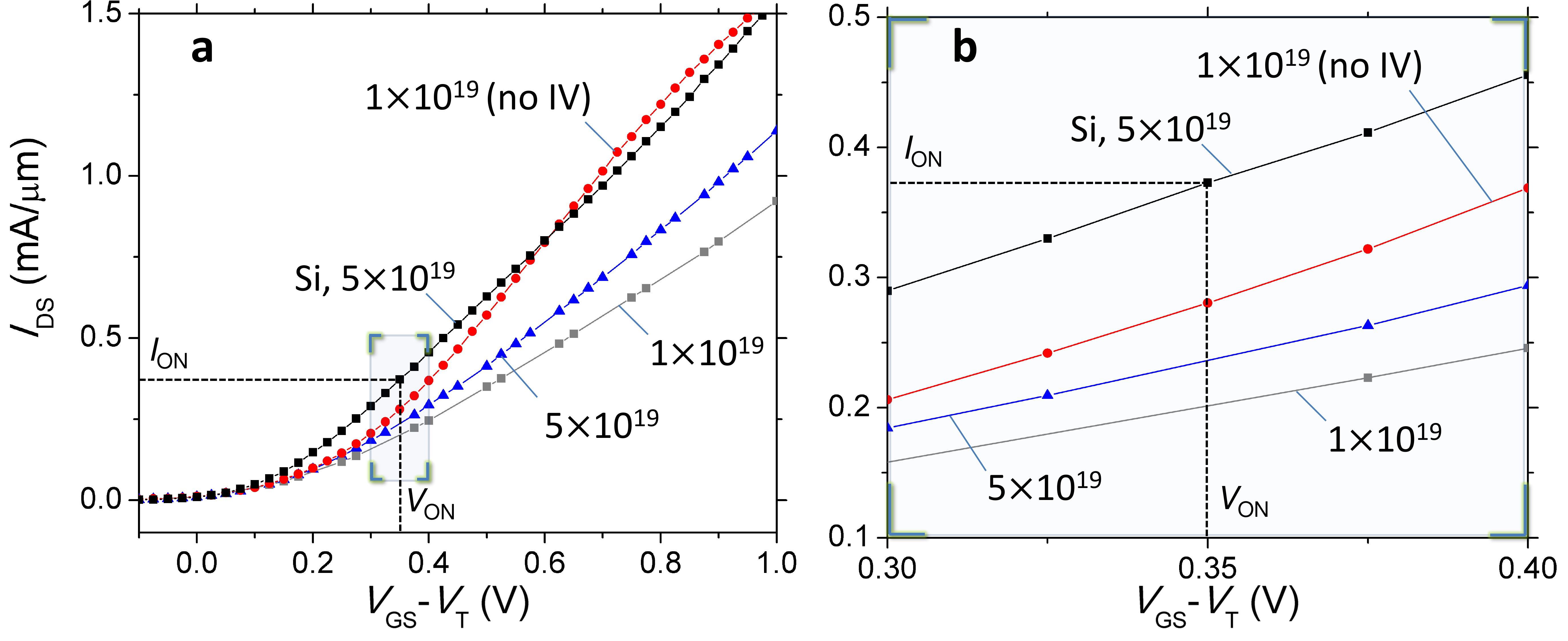}
	\caption{\label{fig:model2}\textbf{ \textbar\; The role of satellite valleys in In$_{\text{0.53}}$Ga$_{\text{0.47}}$As transport.} The individual curves are labeled by their doping density and are In$_{\text{0.53}}$Ga$_{\text{0.47}}$As channels unless noted as Si. \textbf{(a)} With a larger injection velocity, the In$_{\text{0.53}}$Ga$_{\text{0.47}}$As device having no intervalley scattering (no IV) outperforms the device including the peripheral valleys, although it has a smaller quantum capacitance. It also outperforms the In$_{\text{0.53}}$Ga$_{\text{0.47}}$As device having a larger doping density of $5\times10^{19}~\mathrm{cm}^{-3}$ and even crosses-over Si at larger gate biases. \textbf{(b)} Si, however, still has the largest ON-current at the given drive voltage determined by ITRS predictions for future CMOS.} 
\end{figure*}
However, the absence of energetically available peripheral valleys\textemdash or ignoring scattering to available ones\textemdash does have a significant impact on device performance. Transfer curves are shown in Fig.~\ref{fig:model2} for devices with full QCs comparing In$_{\text{0.53}}$Ga$_{\text{0.47}}$As FinFETs with S/D doped to $N_{\mathrm{D}} = 1\times10^{19}~\mathrm{cm}^{-3}$ with intervalley scattering and without (no IV). Transfer curves for Si and In$_{\text{0.53}}$Ga$_{\text{0.47}}$As devices from Fig.~\ref{fig:model} doped to $N_{\mathrm{D}} = 5\times10^{19}~\mathrm{cm}^{-3}$ also are shown for reference. As stated previously, there are competing effects of intervalley scattering on device performance. With intervalley scattering turned on, overall channel injection efficiency is poor ($\gamma  = 36\%$), having significant occupation of the high-scattering rate peripheral valleys, as well as substantial intervalley scattering that, itself, produces back-scattering. In contrast, the quantum capacitance increases with the occupation of the satellite valleys, leading to an overall larger carrier population in the channel for a given gate voltage than otherwise would be found. However, the larger gate control of the device including intervalley scattering is not enough to offset the greater benefits enjoyed by the device where the peripheral valleys are no longer available (Fig.~\ref{fig:model2}, no IV), whose large injection velocity of $4.3\times10^7$~cm/sec is great enough to outperform even the In$_{\text{0.53}}$Ga$_{\text{0.47}}$As device doped to $N_{\mathrm{D}} = 5\times10^{19}~\mathrm{cm}^{-3}$ and included intervalley scattering. This intervalley-scattering-free device even outperforms Si at larger gate biases well above threshold, as its DOS and, thus, quantum capacitance grows with increasing energy. However, at the considered ITRS voltages where the quantum capacitance is lower, it still exhibits a smaller $I_{\mathrm{ON}}$ where there is a more rapid turn-on of the drain current in the Si device just above threshold. Notably, with much higher available doping (e.g., $N_{\mathrm{D}} > 5\times10^{20}~\mathrm{cm}^{-3}$ in Ref.~\onlinecite{liu}), Si devices would be expected to have still better performance relative to III-V devices due to reduced S/D resistances and better contacts. 

Our device results with full quantum corrections can be compared qualitatively to previous studies.  However, quantitative comparison to Si-FinFET experimental results is difficult due to, e.g., different geometries,~\cite{basker,liu} strain considerations, and uncertainty in surface roughness.  The same can be said regarding existing III-V FinFET experimental devices and technologies.~\cite{rado,waldron} It is also difficult to make quantitative comparisons among the experimental results for similar reasons. In addition, our simulations excluded realistic modeling of contact resistance, which was done to isolate the respective channel behavior across materials to study the relevant transport physics considered. For comparison to others simulations, there are no other MC simulators currently that treat the array of quantum effects modeled in this work. However, ballistic quantum transport simulations have exhibited the same qualitative trend exhibited here, namely that Si devices may continue to outperform III-V devices moving forward.~\cite{park,cantley}
\section{Conclusion}
In this study, we provided an ensemble MC methodology with the most complete set of quantum corrections in terms of the number of quantum mechanical effects addressed: far-from-equilibrium degenerate statistics and associated PB of scattering and limited quantum (DOS) capacitance, and confinement effects including altered energy separations between energy valley minima and quantum-confinement-dependent phonon and surface-roughness scattering, in addition to electron redistribution in real space and modified source-to-channel potential barriers. We developed each of our methods individually within this article with a focus on our new contributions and discussed their relevance in terms of nanoscale n-channel FinFET device performance, illustrated through application to example In$_{\text{0.53}}$Ga$_{\text{0.47}}$As and Si devices.

For the treatment of the PB of scattering, we avoid the common use of Fermi-Dirac equilibrium electron distributions to approximate the final state occupation probabilities. Instead, our method directly samples even far-from-equilibrium forward-going and backward-going local electron populations as a function of energy valley and energy, and uses those occupation probabilities self-consistently to model PB. We also introduced sub-carriers (fractional carriers) to suppress classical molecular dynamics carrier-carrier interactions that inherently do not consider the Pauli exclusion principle, with the added benefits of enhancing simulation statistics and minimizing self-forces. Our method of calculating degenerate carrier populations was shown to limit to Fermi-Dirac statistics under equilibrium conditions, while flexibly adapting to more complex distributions under bias.  

We modeled the above-noted quantum-confinement effects via space-, valley-, and orientation-dependent quantum-correction potentials. In doing so, we extended to 3D a treatment of quantum-confined phonon and ionized-impurity scattering developed previously in-house, and found a versatile method for modeling surface-roughness scattering with these potentials that extends to arbitrary potential-well shapes, giving material-, valley-, and orientation-dependent SR scattering in various device geometries.

We showed that collectively these modeled quantum effects can substantially degrade or even eliminate otherwise expected benefits of considered In$_{\text{0.53}}$Ga$_{\text{0.47}}$As devices over industry-standard Si devices, despite lower bulk electron masses, higher mobilities, and higher thermal velocities found in III-V materials, even while neglecting non-ideal contacts and reduced interface quality that are likely to be worse for III-Vs.

We note that it also may be possible to use quantum corrections in the ways described herein within simpler drift-diffusion or hydrodynamic simulators, albeit using more computationally efficient methods for calculating the potentials themselves.
\section{Acknowledgements}
We acknowledge Dharmendar Reddy Palle for aid in software development.  This work was supported in part by a grant from \textit{GLOBALFOUNDRIES USA}, Inc. D.M.C is supported by an NSF graduate fellowship. We thank the Texas Advanced Computing Center (TACC) for generous supercomputing resources and software consultation.
\clearpage
\section{Appendix A: material parameters}
Listed are the simulated band structure and scattering parameters for Si and In$_{\text{0.53}}$Ga$_{\text{0.47}}$As including the lattice constant ($a_{\mathrm{0}}$), mass density ($\rho$), speed of sound ($v_{\mathrm{s}}$), relative dielectric permittivity ($\varepsilon_{\mathrm{r}}$), electron affinity  ($q\mathlarger{\mathlarger{\chi}}$) , non-parabolicity constant ($\alpha$), valley effective mass ($m$), acoustic deformation potential ($\Delta_{\mathrm{ac}}$), deformation field ($DK$), phonon energy ($\hbar\omega$), valley-wise bowing parameter ($C_i$), and intervalley separation ($E_{i\,j}$).
\begingroup
\squeezetable
\begin{table}[h!]
\begin{center}
	\begin{tabular}{  l c c c  }  
	\multicolumn{4}{c}{	}\\
	\hline

	\,						&\quad\quad\quad\quad\quad\quad&	\quad\quad\quad Si\quad\quad\quad					&	\quad\quad Units \quad \quad		\\

	\hline
	$a_{\mathrm{0}}$				&&	\quad5.43$^{\;a}$\quad									&	\quad\AA\quad			\\

	$\rho$						&&	\quad2.33$^{\;a}$\quad									&	\quad g/cm\quad			\\

	$v_{\mathrm{s}}^{l}$			&								&	\quad9.18$^{\;a}$\quad								&	\quad $\times10^{5}$~cm/s\quad			\\

	$v_{\mathrm{s}}^{t}$			&								&	\quad4.70$^{\;a}$\quad								&	\quad $\times10^{5}$~cm/s\quad			\\

	$\varepsilon_{\mathrm{r}}^0$		&									&	\quad11.7$^{\;a}$\quad								&	\quad \bfseries{\---} \quad			\\

	$q\mathlarger{\mathlarger{\mathlarger{\chi}}}$										&&	\quad4.05$^{\;b}$\quad								&	\quad eV \quad			\\

	$\alpha_{\Delta}$												&&	\quad0.5$^{\;c}$\quad								&	\quad eV$^{-1}$ \quad			\\

	$m_{t}^{\Delta}$														&&	\quad0.191$^{\;c}$\quad								&	\quad $m_{\mathrm{e}}$ \quad			\\

	$m_{l}^{\Delta}$														&&	\quad0.983$^{\;c}$\quad								&	\quad $m_{\mathrm{e}}$ \quad			\\

	$\Delta_{\mathrm{ac}}$													&&	\quad5.0 (1.7$^{\;a}$) (9.0$^{\;c}$)\quad								&	\quad eV\quad			\\

	($D_{t}K)_{g1}^{\Delta}\quad$(TA)											&	&	\quad0.40 (0.50$^{\;c}$)\quad								&	\quad $\times10^8$~eV/cm \quad			\\

	$\hbar\omega_{g1}^{\Delta}$													&&	\quad$140^{\;c}$\quad								&	\quad K \quad			\\

	($D_{t}K)_{g2}^{\Delta}\quad$(LA)												&		&	\quad0.64 (0.80$^{\;c}$)\quad								&	\quad $\times10^8$~eV/cm \quad			\\

	$\hbar\omega_{g2}^{\Delta}$														&&	\quad$215^{\;c}$\quad								&	\quad K \quad			\\

	($D_{t}K)_{g3}^{\Delta}\quad$(LO)													&	&	\quad8.73 (11.00$^{\;c}$)\quad								&	\quad $\times10^8$~eV/cm \quad			\\

	$\hbar\omega_{g3}^{\Delta}$													&&	\quad$720^{\;c}$\quad								&	\quad K \quad			\\

	($D_{t}K)_{f1}^{\Delta}\quad$(TA)												&		&	\quad0.24 (0.30$^{\;c}$)\quad								&	\quad $\times10^8$~eV/cm \quad			\\

	$\hbar\omega_{f1}^{\Delta}$														&&	\quad$220^{\;c}$\quad								&	\quad K \quad			\\

	($D_{t}K)_{f2}^{\Delta}\quad$(LA)													&	&	\quad1.59 (2.00$^{\;c}$)\quad								&	\quad $\times10^8$~eV/cm \quad			\\

	$\hbar\omega_{f2}^{\Delta}$														&&	\quad$550^{\;c}$\quad								&	\quad K \quad			\\

	($D_{t}K)_{f3}^{\Delta}\quad$(TO)												&	&	\quad1.59(2.00$^{\;c}$)\quad								&	\quad $\times10^8$~eV/cm \quad			\\

	$\hbar\omega_{f3}^{\Delta}$														&&	\quad$685^{\;c}$\quad								&	\quad K \quad			\\
	\hline
	$^a$Fischetti (1991), Ref.~\onlinecite{fischetti}\\
	$^b$Goldberg (1999), Ref.~\onlinecite{goldberg}\\   
	$^c$Jacoboni (1983), Ref.~\onlinecite{jacoboni2}\\   
\end{tabular}
\end{center}
\end{table}
\endgroup

\clearpage
\begingroup
\squeezetable
\begin{table}[h!]
\begin{center}
	\begin{tabular}{  l c c c c }  
	\multicolumn{5}{c}{	}	\\
	\hline
	\;									&	\quad InAs \quad    				&	\quad GaAs \quad			&	\quad In$_{\text{0.53}}$Ga$_{\text{0.47}}$As \quad					&	\quad\quad Units \quad \quad		\\
	\hline
	$a_{\mathrm{0}}$						&			\quad6.04$^{\;a}$\quad			&	\quad5.64$^{\;a}$\quad				&	\quad5.85\quad												&	\quad\AA\quad			\\

	$\rho$								&		\quad5.67$^{\;a}$					&	\quad5.36$^{\;a}$\quad				&	\quad5.52\quad												&	\quad g/cm\quad			\\

	$v_{\mathrm{s}}^{l}$					&	\quad4.28$^{\;a}$						&	\quad5.24$^{\;a}$					&	\quad4.73\quad												&	\quad $\times10^{5}$~cm/s\quad			\\
	
	$v_{\mathrm{s}}^{t}$					&	\quad2.65$^{\;a}$						&	\quad2.47$^{\;a}$					&	\quad2.57\quad												&	\quad $\times10^{5}$~cm/s\quad			\\

	$\varepsilon_{\mathrm{r}}^0$				&			\quad15.15$^{\;a}$			&	\quad12.90$^{\;a}$				&	\quad14.09												&	\quad \bfseries{\---} \quad			\\

	$\varepsilon_{\mathrm{r}}^\infty$			&	\quad12.75$^{\;a}$					&	\quad10.92$^{\;a}$				&	\quad11.88											&	\quad \bfseries{\---} \quad			\\

	$q\mathlarger{\mathlarger{\mathlarger{\chi}}}$	& 		\quad4.90$^{\;b}$					&	\quad4.07$^{\;b}$\quad						&	\quad4.51													&	\quad eV \quad			\\

	$C_{\Gamma }$						&			\quad\---			&			\quad\---			&		\quad0.48$^{\;g}$									&	\quad eV \quad			\\

	$C_{\mathrm{L}}$						&			\quad\---			&			\quad\---			&		\quad0.58 $(0.33-0.72)^{\;g}$ 									&	\quad eV \quad			\\

	$C_{\mathrm{X}}$						&			\quad\---			&			\quad\---			&		\quad1.09 $(0.08-1.40)^{\;g}$								&	\quad eV \quad			\\

	$E_{\Gamma \mathrm{L}}$						&			\quad0.711$^{\;c}$\quad			&			\quad0.290$^{\;c}$				&		\quad0.487									&	\quad eV \quad			\\

	$E_{\Gamma \mathrm{X}}$						&		\quad1.011$^{\;c}$					&			\quad0.481$^{\;c}$				&				\quad0.610										&	\quad eV \quad			\\

	$(DK)_{\mathrm{po}}$					&		\quad1.06$^{\;a}$					&		\quad13.6 (2.1$^{\;a}$) \quad &			\quad6.95											&	\quad $\times10^8$~eV/cm \quad			\\

	$\hbar\omega_{po}$						&			\quad348$^{\;a}$				&	\quad417$^{\;a}$					&				\quad380										&	\quad K \quad			\\

	$\alpha_{\Gamma}$						&			\quad1.39$^{\;e}$				&	\quad0.69$^{\;f}$		&				\quad1.06										&	\quad eV$^{-1}$ \quad			\\

	$m_{\Gamma}$							&					\quad0.023$^{\;b}$	&	\quad0.067$^{\;d}$				&	\quad0.044												&	\quad \bfseries{\---} \quad		\\

	$\Delta_{\mathrm{ac}}^\Gamma$			&	\quad10.0 (5.8)$^{\;a}$						&	\quad10.0 (5.0)$^{\;a}$					&					\quad10.0								&	\quad eV\quad			\\

	$DK_{(\Gamma\leftrightarrow \mathrm{L})}$				&	\quad8.16 (5.59)$^{\;a}$				& \quad5.25$^{\;a}$					&				\quad6.81										&	\quad $\times10^8$~eV/cm \quad			\\

	$\hbar\omega_{(\Gamma\leftrightarrow \mathrm{L})}$		&	\quad347$^{\;a}$				&		\quad322$^{\;a}$	 		&				\quad335										&	\quad K \quad			\\

	$DK_{(\Gamma\leftrightarrow \mathrm{X})}$				&	\quad8.16 (6.35)$^{\;a}$				& \quad5.28 (5.48)$^{\;a}$					&						\quad6.81							&	\quad $\times10^8$~eV/cm \quad			\\

	$\hbar\omega_{(\Gamma\leftrightarrow\mathrm{X})}$		&	\quad347$^{\;a}$				&	\quad347$^{\;a}$	 			&		\quad347												&	\quad K \quad			\\

	$\alpha_\mathrm{L}$							&			\quad0.54$^{\;e}$				&	\quad0.65$^{\;f}$				&				\quad0.59									&	\quad eV$^{-1}$ \quad			\\

	$m_t^{\mathrm{L}}$							&					\quad0.286$^{\;e}$	&	\quad0.075$^{\;d}$			&	\quad0.187												&	\quad \bfseries{\---} \quad		\\

	$m_l^{\mathrm{L}}$							&					\quad0.286$^{\;e}$	&	\quad1.900$^{\;d}$			&	\quad1.04												&	\quad \bfseries{\---} \quad		\\
	
	$\Delta_{\mathrm{ac}}^{\mathrm{L}}$			&	\quad10.0 (5.8)$^{\;a}$						&	\quad9.2 (5.0)$^{\;a}$				&		\quad9.62												&	\quad eV\quad			\\

	$DK_{(\mathrm{L}\leftrightarrow \mathrm{X})}$						&	\quad4.76 (5.59)$^{\;a}$			& \quad13.6 (5.01)$^{\;a}$				&	\quad8.91													&	\quad $\times10^8$~eV/cm \quad			\\

	$\hbar\omega_{(\mathrm{L}\leftrightarrow \mathrm{X})}$					&	\quad341$^{\;a}$		&	\quad341$^{\;a}$	 		&		\quad341												&	\quad K \quad			\\

	$DK_{(\mathrm{L}\leftrightarrow \mathrm{L})}$						&	\quad5.28 (6.35)$^{\;a}$			& \quad13.6 (5.94)$^{\;a}$				&		\quad9.19											&	\quad $\times10^8$~eV/cm \quad			\\

	$\hbar\omega_{(\mathrm{L}\leftrightarrow \mathrm{L})}$					&\quad338$^{\;a}$			&	\quad338$^{\;a}$			 &		\quad338												&	\quad K \quad			\\

	$\alpha_\mathrm{X}$							&			\quad0.90$^{\;e}$						&	\quad0.36$^{\;f}$		&		\quad0.65												&	\quad eV$^{-1}$ \quad			\\

	$m_t^{\mathrm{X}}$							&					\quad0.640$^{\;e}$				&	\quad0.19$^{\;d}$	&	\quad0.429												&	\quad \bfseries{\---} \quad		\\

	$m_l^{\mathrm{X}}$							&					\quad0.640$^{\;e}$				&	\quad1.9$^{\;d}$	&	\quad1.23													&	\quad \bfseries{\---} \quad		\\
	
	$\Delta_{\mathrm{ac}}^{\mathrm{X}}$			&	\quad10.0 (5.8)$^{\;a}$						&	\quad9.7 (5.0)$^{\;a}$				&		\quad9.86												&	\quad eV\quad			\\

	$DK_{(\mathrm{X}\leftrightarrow \mathrm{X})}$						&	\quad4.76 (3.36)$^{\;a}$		& \quad13.6 (2.99)$^{\;a}$					&		\quad8.91												&	\quad $\times10^8$~eV/cm \quad			\\

	$\hbar\omega_{\mathrm{X}\leftrightarrow \mathrm{X}}$					&	\quad347$^{\;a}$	&	\quad347$^{\;a}$	 			&		\quad347												&	\quad K \quad			\\

	\hline
	$^a$Fischetti (1991), Ref.~\onlinecite{fischetti}\\
	$^b$Goldberg (1999), Ref.~\onlinecite{goldberg}\\
	$^c$Adachi (2009), Ref.~\onlinecite{adachi}\\
	$^d$Blakemore (1982), Ref.~\onlinecite{blakemore}\\
	$^e$Brennan (1984), Ref.~\onlinecite{brennan2}\\
	$^f$Brennan (1988), Ref.~\onlinecite{brennan1}\\
	$^g$Vurgaftman (2001), Ref.~\onlinecite{vurgaftman}\\
\end{tabular}
\end{center}
\end{table}
\endgroup

\bibliography{myBib}{}
\end{document}